\documentclass[journal]{IEEEtran}

\usepackage{amsmath}
\allowdisplaybreaks[4]
\usepackage{amsthm}
\usepackage{amssymb}
\usepackage{verbatim}
\usepackage[linesnumbered,ruled,vlined]{algorithm2e}
\usepackage{bbding}
\usepackage{graphicx}
\usepackage{mathtools}
\usepackage{multirow}
\usepackage{multicol}
\usepackage{booktabs}
\usepackage{makecell}
\usepackage{yhmath}
\usepackage{paralist}
\usepackage{color}
\usepackage{mathrsfs}
\usepackage{xr-hyper}
\usepackage[colorlinks,
 linkcolor=black,
 urlcolor=black,
 anchorcolor=black,
 citecolor=black]{hyperref}
\usepackage{cite}
\usepackage{enumitem}
\usepackage{scalerel}
\usepackage{xcolor,colortbl}
\usepackage{dsfont}
\usepackage{float}
\usepackage{epstopdf}
\usepackage{subfigure}
\usepackage{lipsum}
\usepackage{scalerel}

\begin{document}
\bstctlcite{BSTcontrol}

\title{Cache-Aware Cooperative Multicast Beamforming in Dynamic Satellite-Terrestrial Networks}

\author{{Shuo~Yuan,~\IEEEmembership{Member,~IEEE},
			Yaohua~Sun,
			Mugen~Peng,~\IEEEmembership{Fellow,~IEEE}}
	\thanks{Copyright \copyright 2024 IEEE. Personal use of this material is permitted. However, permission to use this material for any other purposes must be obtained from the IEEE by sending a request to pubs-permissions@ieee.org. This work was supported in part by the National Key Research and Development Program of China under Grant 2022YFB2902600, in part by the National Natural Science Foundation of China under Grants 62371071, and in part by the Young Elite Scientists Sponsorship Program by the China Association for Science and Technology under Grant 2021QNRC001. (\emph{Corresponding author: Yaohua Sun.})}
	\thanks{The authors are with the State Key Laboratory of Networking and Switching Technology, Beijing University of Posts and Telecommunications, Beijing 100876, China (e-mail: yuanshuo@bupt.edu.cn; sunyaohua@bupt.edu.cn; pmg@bupt.edu.cn). }
}

\maketitle

\begin{abstract}
	With the burgeoning demand for data-intensive services, satellite-terrestrial networks (STNs) face increasing backhaul link congestion, deteriorating user quality of service (QoS), and escalating power consumption.
	Cache-aided STNs are acknowledged as a promising paradigm for accelerating content delivery to users and alleviating the load of backhaul links.
	However, the dynamic nature of low earth orbit (LEO) satellites and the complex interference among satellite beams and terrestrial base stations pose challenges in effectively managing limited edge resources.
	To address these issues, this paper proposes a method for dynamically scheduling caching and communication resources, aiming to reduce network costs in terms of transmission power consumption and backhaul traffic, while meeting user QoS demands and resource constraints.
	We formulate a mixed timescale problem to jointly optimize cache placement, LEO satellite beam direction, and cooperative multicast beamforming among satellite beams and base stations.
	To tackle this intricate problem, we propose a two-stage solution framework, where the primary problem is decoupled into a short-term content delivery subproblem and a long-term cache placement subproblem.
	The former subproblem is solved by designing an alternating optimization approach with whale optimization and successive convex approximation methods according to the cache placement state, while cache content in STNs is updated using an iterative algorithm that utilizes historical information.
	Simulation results demonstrate the effectiveness of our proposed algorithms, showcasing their convergence and significantly reducing transmission power consumption and backhaul traffic by up to 52\%.
\end{abstract}


\IEEEpeerreviewmaketitle

\section{Introduction}

\IEEEPARstart{W}{ith} the rapid development of the Internet of Things (IoT) and mobile communication technologies, connections are extending across numerous fields, including industry, transportation, and agriculture, facilitating interactions among humans, machines, and things \cite{zhu2022integratedsatelliteterrestrial}.
The number of connected devices is predicted to exceed 24.1 billion by 2030 \cite{zhang2021auctionbasedmultichannel}.
In this context, there is an increasingly urgent need to ensure ubiquitous coverage for such massive and widely distributed devices, rather than solely focusing on improving transmission quality in limited areas \cite{chai2023jointmultitask,zhou2021machinelearningbased,lin2021supportingiot,liu2024jointuser}.
Consequently, the integration of low earth orbit (LEO) satellite networks with terrestrial networks is emerging as a crucial solution and drawing significant attention due to its capacities of seamless coverage and high-throughput data transmission \cite{yuan2023softwaredefined,zhu2022creatingefficient,yuan2024jointnetwork}.
Meanwhile, the dramatic increase in connected devices and data-intensive applications is expected to significantly escalate the demands for mobile data traffic.
International Telecommunication Union Radiocommunication Sector estimates an explosive growth in global mobile traffic per month to 2194 exabytes (EB) in 2028 and 5016 EB in 2030 \cite{itu-r2015imttraffic}, while Ericsson forecasts that video content poised to constitute 80\% of all mobile data traffic by 2028 \cite{zhang2024modelinganalysis}.
These trends present major challenges to the limited resources and backhaul capacities of satellite-terrestrial networks (STNs).

To address these challenges in STNs, edge caching emerges as a beneficial solution.
It involves storing popular content on edge nodes, such as satellites and base stations, enabling direct content delivery to users without relying on cloud servers, thereby significantly reducing backhaul traffic \cite{liu2022deeplearningenabled,ngo2022twotiercacheaided,zhang2018cooperativeedge}.
However, the challenge in cache-aided STNs lies in the dynamic accessibility of edge nodes, like satellites, whose rapid movements can render cached content temporarily unreachable, potentially leading to a poor user experience.
Therefore, effective cache placement within the STN requires careful consideration of not only which content needs to be cached and at which edge node but also when it should be cached, taking into account the movement of satellites.

Beyond cache placement, the efficiency of content delivery plays a pivotal role in enhancing system performance within cache-aided networks \cite{lin2023userfeaturesaware,tao2016contentcentricsparse,sun2020qoedriventransmissionaware,wu2020jointlongterm}.
During the content delivery stage, considering the concurrent request for popular content among multiple users, coupled with the advantages of broadcast transmission and the extensive coverage capabilities offered by satellite communications, the adoption of multicast beamforming emerges as a feasible and effective content delivery strategy within STNs \cite{han2022jointcache,khammassi2024precodinghighthroughput}.
Multicast beamforming allows for meeting the same requests from multiple mobile users on the same resource block, thus avoiding repetitive transmissions and enhancing spectral efficiency.
In addition, clustering satellites and base stations to form specific groups for content delivery facilitates satellite-terrestrial cooperative transmission.
This approach not only improves the signal-to-interference-plus-noise ratio (SINR) for users but also further reduces both backhaul load and resources consumed.

\subsection{Related Works}
\subsubsection{Satellite-Assisted Caching}
As satellite networks are increasingly recognized as key components of beyond-5G and 6G networks, satellite-assisted caching has emerged as a promising approach to enhance network throughput and quality of service, as evidenced by numerous studies \cite{zhu2022cooperativemultilayer,liu2018distributedcaching,hu2023jointoptimization,zhang2023jointoptimization,li2023multiagentdrl}.
In \cite{zhu2022cooperativemultilayer}, a three-layer cooperative caching model in STNs is introduced, which involves the base station, satellite, and ground gateway working together to cache popular content and provide services to users.
The authors derive the cache hit probability for different caching locations and propose an alternating iterative algorithm to optimize cache placement, aiming to minimize the average content retrieval latency for users.
In \cite{liu2018distributedcaching}, the authors formulate a distributed cache placement optimization problem for LEO satellite networks, and an exchange-stable matching algorithm based on a many-to-many matching game with externalities is proposed to minimize the content access latency.
To minimize both service latency and energy consumption, the authors of \cite{hu2023jointoptimization} formulate a cache placement optimization problem for STNs based on cached data density, with constraints including constellation structure and cache hit probability, and design an iterative algorithm that utilizes a reinforcement learning method and the water-filling algorithm.
In \cite{zhang2023jointoptimization}, the authors focus on improving system performance and reducing power consumption in STNs, where courtship movements and random flights of the mayflies-based algorithm is proposed to jointly optimize the cache placement and power allocation under the limitations of cache capacity and power resource.
To maximize energy efficiency, a joint optimization problem of user association, power control, and cache placement in STNs is formulated in \cite{li2023multiagentdrl}, and then a multiagent deep deterministic policy gradient algorithm is proposed.

However, most of the mentioned studies primarily concentrate on cache placement within STNs, the optimization for content delivery is often not as thoroughly explored.
Moreover, the capabilities of satellite communications, such as broadcast transmission and wide coverage, are not fully underexploited in this research area.

\subsubsection{Multicast Beamforming for Content Delivery}
To enhance the efficiency of content delivery, a series of studies have investigated the multicast beamforming in STNs.
Compared to traditional point-to-point unicast beamforming, multicast beamforming is well-suited for content delivery and holds great potential in various applications, including mobile multimedia, software downloads and updates, as well as virtual reality \cite{khammassi2024precodinghighthroughput}.
In \cite{zhu2018cooperativemultigroup}, the downlink cooperative multi-group multicast transmission in STNs is investigated, where satellite and base stations provide the multicast service for users cooperatively, and an iterative algorithm is proposed to optimize the beamforming vectors of the satellite and base stations to minimize the total power consumption.
The authors of \cite{xiao2023multigroupmulticast} investigate multi-group multicast beamforming for geostationary earth orbit (GEO) satellite communications, where the optimization objective is to maximize the energy efficiency and the constraints including power consumption outage and total power usage, and then a low-complexity algorithm based on semidefinite relaxation is proposed.
To obtain the max-min fair system capacity, the authors of \cite{ma2024resourcescheduling} focus on the downlink multi-group multicast transmission in an ultra-dense satellite network, where satellites serve as aerial base stations to cooperatively provide multicast service for users, and an alternating iterative algorithm based on many-to-many matching model and successive convex approximation (SCA) method is proposed to jointly optimize downlink beamforming and subchannel assignment.

Furthermore, several studies focus on integrating cache placement and multicast beamforming in STNs.
In \cite{han2022jointcache}, a joint optimization problem is formulated that encompasses cache placement, satellite and base station clustering, and multicast beamforming, and a two-step algorithm based on a penalty concave-convex procedure is proposed to maximize the network throughput.
In \cite{wang2020overlaycoded}, the authors integrate network coding with multicast and cache placement in STNs to enhance network throughput and propose a tree-structured multicast group merging strategy followed by a solution based on Dijkstra's algorithm.

\subsection{Motivations and Contributions}
In the cache-aided STN, the limited onboard resources of satellites, including cache storage, power, and radio frequency resources, highlight the importance of developing effective resource management strategies for cache placement and content delivery to improve resource utilization and reduce network cost.
However, the aforementioned studies \cite{zhu2022cooperativemultilayer, liu2018distributedcaching, hu2023jointoptimization, zhang2023jointoptimization, li2023multiagentdrl} focus primarily on optimizing cache placement, often overlooking the potential benefits of satellite communication, such as broadcasting and wide coverage, in the content delivery stage.
On the other hand, the studies \cite{zhu2018cooperativemultigroup, xiao2023multigroupmulticast, ma2024resourcescheduling}, which explore multicast beamforming in STNs for improved network throughput and energy efficiency, fail to consider the effects of full cloud-based content delivery on quality of service and the load of backhaul links.
Moreover, the dynamic nature of STN topologies, particularly due to the rapid movement of satellites, is not adequately addressed in these studies \cite{zhu2022cooperativemultilayer, liu2018distributedcaching, hu2023jointoptimization, zhang2023jointoptimization, li2023multiagentdrl, zhu2018cooperativemultigroup, xiao2023multigroupmulticast, ma2024resourcescheduling}, potentially affecting the effectiveness of their models for dynamic STNs.
While the authors in \cite{han2022jointcache} and \cite{wang2020overlaycoded} consider both cache placement and multicast beamforming, they do not examine the integration of satellite beam scheduling with multicast beamforming in dynamic cache-aided STNs, nor do they focus on optimizing energy efficiency.

Motivated by the aforementioned observations, this work investigates a dynamic cache-aided STN in which satellites and terrestrial base stations cooperatively multicast beamforming to provide content-centric data services to users.
Specifically, we focus on a two-timescale optimization problem involving cache placement, beam direction control, and cooperative multicast beamforming within a dynamic STN, aiming to minimize transmission power consumption and backhaul traffic costs.
In this scenario, the cache placement strategy is updated on a long-term basis, while the beam direction control and multicast beamforming strategies are optimized more frequently on a short-term basis.
To deal with the problem, we propose an alternating optimization-based solution, which introduces the whale optimization algorithm (WOA) to optimize the beam direction and an SCA-based algorithm for cooperative multicast beamforming, as well as an iterative algorithm, which utilizes historical information for the long-term cache placement optimization.
The main contributions of this paper are concluded as follows.

\begin{enumerate}
	\item
	      A two-timescale problem of joint satellite beam direction control, multicast beamforming, and cache placement is formulated in the context of a dynamic multi-beam satellite-terrestrial network for the efficient delivery of content-centric data services to users.
	      The objective is to minimize the overall network cost, considering transmission power consumption and backhaul traffic, while satisfying the quality of service demands for each multicast group and adhering to resource constraints.
	      Specifically, the cached contents available in satellites and base stations are periodically updated on a long-term basis based on user content requests and the wireless coverage status, whether provided by satellite beams or base stations.
	      Moreover, to enhance the efficiency of content delivery and alleviate interference among satellite beams and base stations, beam direction control and multicast beamforming are optimized more frequently on a short-term basis according to the cache placement state.
	\item
	      To tackle this mixed timescale optimization problem, we propose a two-stage solution framework addressing both long-term cache placement and short-term content delivery.
	      Specifically, we introduce an alternating optimization-based approach, consisting of an improved whale optimization algorithm (WOA) for beam direction control and a successive convex approximation (SCA)-based algorithm for cooperative multicast beamforming, to solve the short-term content delivery subproblem.
	      In addition, we devise an iterative algorithm that utilizes historical user content requests and channel state information for the long-term cache placement subproblem.
	\item
	      The impacts of critical parameters such as cache capacities and the trade-off factor between power consumption and backhaul traffic on the performance of network cost are evaluated, showcasing the superiority of our proposed algorithm over different baseline solutions.
	      Results demonstrate the convergence of our proposed algorithms, achieving significant reductions in transmission power consumption and backhaul traffic by up to 52\%.
\end{enumerate}

\subsection{Organization}
The rest of this paper is organized as follows.
Section \ref{sec:systemModel} presents the system model for cooperative multicast beamforming in a cache-aided STN.
In Section \ref{sec:problemFormulation}, the mixed-timescale caching and multicast beamforming optimization problem is formulated, and the primal problem is decomposed into two-stage subproblems.
Section \ref{sec:sol4shortTerm} proposes a solution for the short-term problem of joint satellite beam direction control, access points clustering, and cooperative multicast beamforming, while Section \ref{sec:sol4longTerm} presents a solution for the long-term cache placement problem.
Simulation results are discussed in Section \ref{sec:simulation}, followed by the conclusions in Section \ref{sec:conclusions}.

\begin{figure}[t]
	\centering
	\includegraphics[width=0.485\textwidth]{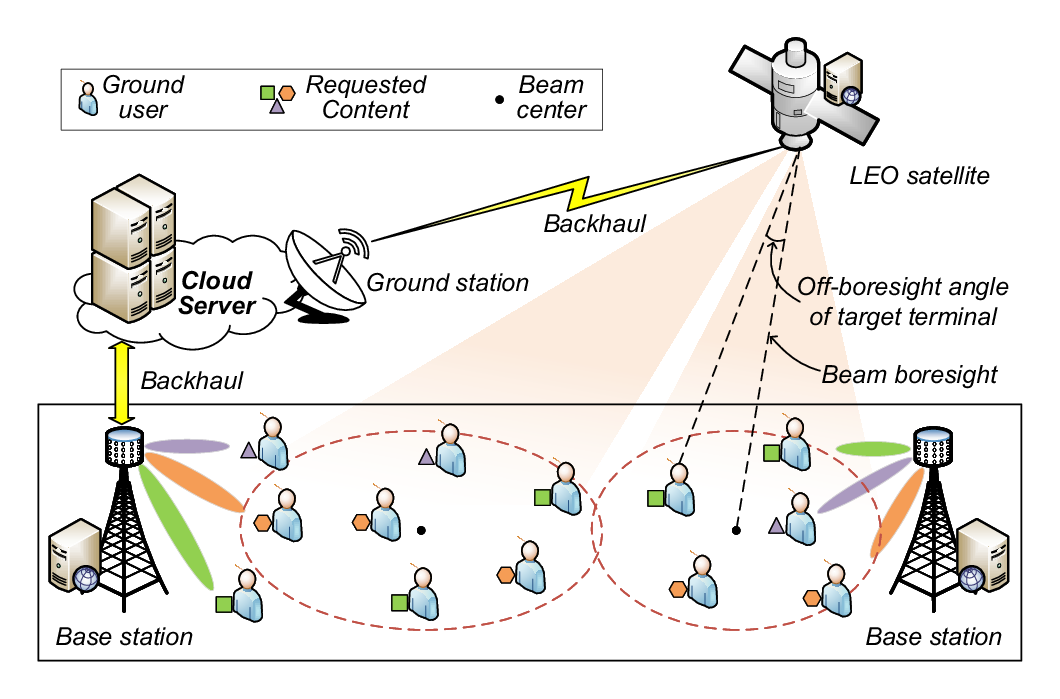}
	\caption{Illustration of the studied satellite-terrestrial network topology.}
	\label{fig:NETtopology}
\end{figure}

\section{System Model}
\label{sec:systemModel}
As shown in Fig. \ref{fig:NETtopology}, we consider a content-centric service provisioning scenario in cache-aided satellite-terrestrial networks, where one multi-beam satellite equipped with $K_s$ feeds generates $M$ adjacent beams, $\mathcal{M}=\{1, \ldots, M\}$, and collaborates with $L$ terrestrial base stations (BSs), $\mathcal{L}=\{1, \ldots, L\}$, each equipped with $K_b$ antennas, to deliver contents to a set of single-antenna ground users, $\mathcal{N}=\{1, \ldots, N\}$.
A library of content files, $\mathcal{F}=\{1, \ldots, F\}$, is stored in the cloud server, with a portion cached at satellite or BSs, awaiting delivery to the users.
The network operation period is discretized into $T$ equal-time slots, indexed by $t\in \mathcal{T} = \{1,\ldots,T \}$.
Let $\pi_f (t)$ be the multicast user group consisting of all the users who requested $f \in \mathcal{F}$ at time slot $t$.
Assuming that each user requests a single content item per time slot, we have $\pi_i(t) \cap \pi_j (t)=\emptyset$ for any distinct $i, j \in \mathcal{F}$, and the total number of requested content items across all users does not exceed $N$, denoted as $\sum_{f \in \mathcal{F}} |\pi_f(t) | \leq N$, for each time slot $t \in \mathcal{T}$.
Each user group, e.g., $\pi_{f} (t)$, is served by an AP cluster $\{\mathcal{L} _f\left( t \right) ,\mathcal{M} _f\left( t \right) \}$, which comprises a set of BSs $\mathcal{L} _f\left( t \right)$ and a set of satellite beams $\mathcal{M} _f\left( t \right)$, through cooperative multicast transmission.
The BSs and satellite have limited cache capacity and are connected to the cloud server via high-speed backhaul.
Each BS and satellite beam in a cluster retrieves the requested contents either from its local cache or from the cloud server through the backhaul link.

\subsection{Communication Model}

Let $\mathbf{c}_{m}^{SB}(t)$ denotes the Earth-Centered Earth-Fixed (ECEF) coordinate of the beam center of satellite beam $m$ on the surface of Earth at time slot $t$, and $\mathbf{c}_{n}^{U}(t)$ represents the ECEF coordinate of ground user $n$.
For the multi-beam satellite, the channel vector from satellite beam $m$ to user $n$ at time slot $t$ can be expressed as \cite{zheng2012genericoptimization,wang2022robustbeamforming}
\begin{equation}
	\mathbf{h}_{m,n}(t)=\left( G_n\mathbf{a}\left( \mathbf{c}_{m}^{SB}\left( t \right) ,\mathbf{c}_{n}^{U}\left( t \right) \right) \right) ^{\frac{1}{2}}\odot \boldsymbol{\nu }_{m,n}^{(t)},
\end{equation}
where ${G_{n}}$ is the receiving antenna gain of user $n$ and $\boldsymbol{\nu }_{m,n}^{(t)}$ denotes the channel response vector.
$\mathbf{a}\left( \mathbf{c}_{m}^{SB}\left( t \right) ,\mathbf{c}_{n}^{U}\left( t \right) \right) $, conveniently simplified as $\mathbf{a}_{m,n}^{(t)}$, represents the satellite beam gain vector, which is a $K_s$-dimensional satellite beam gain vector.
Its $k_s$-th element from satellite antenna feed $k_s$ to user $n$, denoted as $a_{m,n}^{k_s,(t)}$, can be approximated as \cite{wang2022robustbeamforming,yuan2024jointbeam}
\begin{equation}
	\label{eq:channelTransGain}
	a_{m,n}^{k_s,(t)}=G_{SB}^{\max} \left( \frac{J_1\left( \mu \right)}{2\mu}+36\frac{J_3\left( \mu \right)}{\mu ^3} \right)^2,
\end{equation}
where $G_{SB}^{\max}$ is the maximal gain of the satellite beam, and $J_1\left( \cdot \right) $ and $J_3\left( \cdot \right) $ are the first and third order Bessel functions, respectively.
The factor $\mu$ is given by $\mu =2.07123\sin ( \alpha _{m,n}^{( t )} ) /\sin ( \alpha _{m,3\mathrm{dB}}^{( t )} ) $, where $\alpha _{m,3\mathrm{dB}}^{( t )}$ denotes the $3 \rm{dB}$ beamwidth angle of satellite beam $m$, and $\alpha _{m,n}^{\left( t \right)}$ represents the off-boresight angle, which is the angular difference between the direction of satellite beam $m$ and the direction toward user $n$, as illustrated in Fig. \ref{fig:NETtopology}.

For channel response vector ${\boldsymbol {\nu }_{m,n}^{(t)}}$, its $k_s$-th element ${\nu}_{m,n}^{k_s,(t)}$ towards satellite antenna feed $k_s$ can be written as \cite{lin2021supportingiot}
\begin{equation}
	\nu _{m,n}^{k_s,(t)}= \frac{\epsilon}{4\pi d_{m,n}^{\left( t \right)}f_c}e^{-j\frac{2\pi f_c}{\epsilon}d_{m,n}^{\left( t \right)}},
\end{equation}
where $\epsilon$ is the speed of light, $f_c$ is the carrier frequency, and $d_{m,n}^{\left( t \right)}$ denotes the distance between the satellite to which beam $m$ belongs and user $n$ at time slot $t$.

The terrestrial channel between base stations and users is modeled by the Rayleigh channel with shadowing effect \cite{han2022jointcache}.
Similarly, the channel vector between base station $l$ and user $n$ at time slot $t$ is $\mathbf{u}_{l, n}(t)$.
Meanwhile, the downlink beamforming vector from satellite beam $m$ and base station $l$ to group $\pi_{f}(t)$ are denoted as $\mathbf{w}_{m,f}(t)\in \mathbb{C} ^{K_s\times 1}$ and $\mathbf{v}_{l, f}(t) \in \mathbb{C} ^{K_b\times 1}$, respectively.
Subsequently, the signal received at user $n$ in multicast group $\pi_{f}(t)$ can be expressed as
\begin{equation}
	\begin{aligned}
		y_n =
		 & {{\underbrace{
							\sum_{m\in \mathcal{M}}{\mathbf{h}_{m,n}(t)^H\mathbf{w}_{m,f}(t)s_f(t)}+\sum_{l\in \mathcal{L}}{\mathbf{u}_{l,n}(t)^H\mathbf{v}_{l,f}(t)s_f(t)}
		}_{\text{desired signal}}}}              \\
		 & +
		{{\underbrace{
					\sum_{j\in \{\mathcal{F} \backslash f\}}{\sum_{m\in \mathcal{M}}{\mathbf{h}_{m,n}(t)^H\mathbf{w}_{m,j}(t)s_j(t)}}
		}_{\text{satellite beam interference}}}} \\
		 & + {{\underbrace{
					\sum_{j\in \{\mathcal{F} \backslash f\}}{\sum_{l\in \mathcal{L}}{\mathbf{u}_{l,n}(t)^H\mathbf{v}_{l,j}(t)s_j(t)}}
				}_{\text{base station interference}}}}
		+
		{{\underbrace{\vphantom{ \sum_{j\in \{ \mathcal{F} \backslash f \}} }
					z_n( t )
				}_{\mathclap{\text{noise}}}}},
	\end{aligned}
\end{equation}
where $s_f(t)$ is the data symbol of the content requested by group $\pi_{f}(t)$ with $\mathbb{E}[|s_f(t)|^2]=1$, and $z_n(t) \sim \mathcal{C N}\left(0, \sigma_n^2\right)$ is the additive white Gaussian noise at user $n$.

The received SINR for user $n \in \pi_f(t)$ is expressed as
\begin{equation}
	\begin{aligned}
		\Gamma _n(t)=\frac{|\mathbf{h}_n(t)^H\mathbf{w}_f(t)+\mathbf{u}_n(t)^H\mathbf{v}_f(t)|^2}{\sum\limits_{j\in \{\mathcal{F} \backslash f\}}{|\mathbf{h}_n(t)^H\mathbf{w}_j(t)+\mathbf{u}_n(t)^H\mathbf{v}_j(t)|^2}+\sigma _{n}^{2}},
	\end{aligned}
\end{equation}
where $\mathbf{h}_n(t)=[ \mathbf{h}_{1,n}(t)^{H},\mathbf{h}_{2,n}(t)^{H},...,\mathbf{h}_{M,n}(t)^{H}] ^H\in \mathbb{C} ^{MK_s\times 1}$ and $\mathbf{u}_n( t ) =[ \mathbf{u}_{1,n}( t ) ^H,\mathbf{u}_{2,n}( t ) ^H,...,\mathbf{u}_{L,n}( t ) ^H ] ^H\in \mathbb{C} ^{LK_b\times 1}$ denote the aggregate channel vectors to ground user $n$ from all satellite beams and base stations at time slot $t$, respectively.
$\mathbf{w}_f( t ) =[ \mathbf{w}_{1,f}( t ) ^H,\mathbf{w}_{2,f}( t ) ^H,...,\mathbf{w}_{M,f}( t ) ^H ] ^H\in \mathbb{C} ^{MK_s\times 1}$ and $\mathbf{v}_f( t ) =[ \mathbf{v}_{1,f}( t ) ^H,\mathbf{v}_{2,f}( t ) ^H,...,\mathbf{v}_{L,f}( t ) ^H ] ^H\in \mathbb{C} ^{LK_b\times 1}$ denote the aggregate beamforming vectors to group $\pi_f(t)$ from all satellite beams and base stations, respectively.

To ensure successful decoding of the transmitted message, for any user $n$ in group $\pi_f(t)$, the SINR should meet $\Gamma_n (t) \geq \gamma_f$, where $\gamma_f$ represents the minimum SINR required by the users in group $\pi_f(t)$.
Hence, the minimum SINR vector is defined as $\boldsymbol{\gamma }=\left[ \gamma _{1},\gamma _{2},...,\gamma _{F} \right] $.

\subsection{Cache Model}

A binary vector $\mathbf{q}_{f} \left( t \right)^{H} \in \left\{0,1\right\}^{(L+1) \times 1}$ is introduced to characterize the cache placement strategy of the content $f \in \mathcal{F}$ at the satellites and the base station at time slot $t$, which is updated every $T$ time slots, e.g., $t\in \left\{ \theta T|\theta \in \mathbb{Z} + \right\} $.
The cache placement strategy of all content for all APs at time slot $t$ is represented by $\mathbf{q}\left( t \right) \in \{0,1\}^{F\times (L+1)}$.
$q_{f, 0} \left( t \right)=1$ indicates that content $f$ is cached in the satellite at time slot $t$, with 0 indicating it is not cached.
For base stations, $q_{f, l}(t)$ for all $l > 0$ denotes whether content $f$ is cached in base station $l$ at time slot $t$, where 1 means it is cached and 0 means it is not.
For each BS $l \in \mathcal{L}$ or satellite beam $m \in \mathcal{M}$, content $f$ can be directly accessed from its local storage if it has been previously cached.
Otherwise, it must be retrieved from the cloud server via the backhaul link.

Considering the limited caching resources at APs, the cache placement strategy should satisfy the following constraint
\begin{equation}
	\label{eq:cachingStorage}
	\sum_{f\in \mathcal{F}}{q_{f,i}\left( t \right)}\le \phi _i,\forall i\in \left\{ 0,\mathcal{L} \right\} ,t\in \mathcal{T} ,
\end{equation}
where $ \phi _0$ and $ \phi _i, \forall i \in \mathcal{L}$ denote the available local storage size of the satellite and base station $i$, respectively.

\subsection{AP Clustering}

The matrix $\mathbf{e}(t)$ is introduced to characterize the clustering of APs for all user groups at time slot $t$, which is defined as
\begin{equation}
	\label{eq:APclusteringMatrix}
	\mathbf{e}(t)=[\mathbf{e}^\prime(t),\mathbf{e}^{\prime\prime}(t)] \in \left\{ 0,1 \right\} ^{F\times \left( M+L \right)} ,
\end{equation}
where
\begin{equation}
	\begin{aligned}
		 & \mathbf{e}^{\prime}(t)=\left[ \mathbf{e}_{1}^{\prime}\left( t \right) ,\ldots,\mathbf{e}_{M}^{\prime}\left( t \right) \right] \in \left\{ 0,1 \right\} ^{F\times M},                                 \\
		 & \mathbf{e}^{\prime\prime}\left( t \right) =\left[ \mathbf{e}_{1}^{\prime\prime}\left( t \right) ,\ldots,\mathbf{e}_{L}^{\prime\prime}\left( t \right) \right] \in \left\{ 0,1 \right\} ^{F\times L}.
	\end{aligned}
\end{equation}
Here, $\mathbf{e}_{m}^{\prime}(t)=\left[ e_{1,m}^{\prime}\left( t \right) ,...,e_{F,m}^{\prime}\left( t \right) \right] ^H$ and $\mathbf{e}_{l}^{\prime\prime}\left( t \right) =\left[ e_{1,l}^{\prime\prime}\left( t \right) ,...,e_{F,l}^{\prime\prime}\left( t \right) \right] ^H$ represent the clustering status of satellite beam $m$ and base station $l$ to serve all $F$ user groups at time slot $t$.
Elements $e_{f, m}^{\prime}(t)=1$ and $e_{f, l}^{\prime \prime}(t)=1$ indicate that satellite beam $m$ and base station $l$ are scheduled to serve multicast group $\pi_f(t)$, respectively, and $e_{f, m}^{\prime}(t)=0$ and $e_{f, l}^{\prime \prime}(t)=0$ otherwise.

In addition, if $e_{f, m}^{\prime}(t) = 0$ or $e_{f, l}^{\prime \prime}(t) = 0$, then the beamforming vector from satellite beam $m$ or from base station $l$ to multicast group $\pi_f(t)$ is zero.
This indicates that the following constraints must be satisfied
\begin{equation}
	\label{eq:beamformingConst}
	\begin{aligned}
		 & \left( 1-e_{f,m}^{\prime}\left( t \right) \right) \mathbf{w}_{m,f}\left( t \right) =\mathbf{0},\forall m\in \mathcal{M} ,t\in \mathcal{T} ,       \\
		 & \left( 1-e_{f,l}^{\prime\prime}\left( t \right) \right) \mathbf{v}_{l,f}\left( t \right) =\mathbf{0},\forall l\in \mathcal{L} ,t\in \mathcal{T} .
	\end{aligned}
\end{equation}

\subsection{Cost Model}

In this paper, we focus on the total network cost for content delivery, which encompasses both the backhaul traffic cost and the transmission power consumption.
Due to the limited power resources at the base stations and on the satellite, the following constraints must be satisfied by all BSs, satellite beams, and the satellite itself.
\begin{subequations}
	\label{eq:powerConst}
	\begin{align}
		\begin{split}
			\label{eq:powerConstBS}
			& \sum_{f\in \mathcal{F}}{\left\| \mathbf{v}_{l,f}\left( t \right) \right\| _{2}^{2}}\le P_{B}^{\max},\forall l\in \mathcal{L} ,t\in \mathcal{T} ,
		\end{split}
		\\
		\begin{split}
			\label{eq:powerConstSB}
			& \sum_{f\in \mathcal{F}}{\left\| \mathbf{w}_{m,f}\left( t \right) \right\| _{2}^{2}}\le P_{SB}^{\max},\forall m\in \mathcal{M} ,t\in \mathcal{T} ,
		\end{split}
		\\
		\begin{split}
			\label{eq:powerConstS}
			& \sum_{m\in \mathcal{M}}{\sum_{f\in \mathcal{F}}{\left\| \mathbf{w}_{m,f}\left( t \right) \right\| _{2}^{2}}}\le P_{S}^{\max},\forall t\in \mathcal{T} ,
		\end{split}
	\end{align}
\end{subequations}
where $P_B^{\max }$, $P_{SB}^{\max}$, and $P_S^{\max }$ are the maximum transmission powers of BSs, satellite beams, and the satellite, respectively.
With these power constraints, the total transmission power consumption by all the BSs and the satellite at time slot $t$ is defined as
\begin{equation}
	\label{eq:powerCost}
	C_P(t)=\sum_{l\in \mathcal{L}}{\sum_{f\in \mathcal{F}}{\left\| \mathbf{v}_{l,f}\left( t \right) \right\| _{2}^{2}}}+\sum_{m\in \mathcal{M}}{\sum_{f\in \mathcal{F}}{\left\| \mathbf{w}_{m,f}\left( t \right) \right\| _{2}^{2}}}.
\end{equation}

In our considered multicast communication scenario, fixed-rate transmission \cite{dai2013sparsebeamforming} is employed, where the data rate of retrieving requested content from the cloud server must be at least as large as the transmission rate from the AP to the users \cite{tao2016contentcentricsparse}.
Therefore, the transmission rate $R_f(t)$ achieved by group $\pi _f\left( t \right) $ is expressed as
\begin{equation}
	R_f(t)=B \log _2\left(1+\gamma_f\right),
\end{equation}
where $B$ is the communication bandwidth.
Then, we can present the total backhaul traffic cost of the network at time slot $t$ as
\begin{equation}
	\label{eq:backhaulCost}
	\begin{aligned}
		C_B(t)
		 & =\sum_{l\in \mathcal{L}}{\sum_{f\in \mathcal{F}}{e_{f,l}^{\prime}\left( t \right)}}\left( 1-q_{f,l}\left( t \right) \right) R_f\left( t \right)          \\
		 & +\sum_{m\in \mathcal{M}}{\sum_{f\in \mathcal{F}}{e_{f,m}^{\prime\prime}}}\left( t \right) \left( 1-q_{f,0}\left( t \right) \right) R_f\left( t \right) .
	\end{aligned}
\end{equation}

Consequently, the total network cost over a given period is presented as
\begin{equation}
	C=\sum_{t\in \mathcal{T}}{\left( C_P\left( t \right) + \rho C_B\left( t \right) \right)},
\end{equation}
where $\rho > 0$ serves as a weighting factor that determines the trade-off between backhaul traffic cost and transmission power consumption.

\section{Problem Formulation and Decomposition}
\label{sec:problemFormulation}

In this section, we first formulate the mixed timescale optimization problem, which jointly addresses cache placement, beam direction control, AP clustering, and multicast beamforming.
Next, we decompose the primal problem into two subproblems based on different timescales.
This decomposition leads to a two-stage solution framework, which lays the foundation for further algorithm design.

\subsection{Problem Formulation}

Our objective is to minimize the total network cost by optimizing the cache placement $\mathbf{q}=\left\{ \mathbf{q}\left( t \right) |t\in \left\{ \theta T|\theta \in \mathbb{Z}^{+} \right\} \right\}$ every $T$ time slots, and the content delivery strategy $\mathbf{\Upsilon }\left( t \right)  =\left\{ \mathbf{c}^{SB}(t),\mathbf{e}\left( t \right) ,\mathbf{w}\left( t \right) ,\mathbf{v}\left( t \right) \right\}$ at each time slot $t$, which including the satellite beam center $\mathbf{c}^{SB}(t)=\{\mathbf{c}_{m}^{SB}(t)|m\in \mathcal{M} \}$, BS and satellite beam clustering $\mathbf{e}\left( t \right)$, as well as multicast beamforming $\mathbf{w}\left( t \right)  = \left\{ \mathbf{w}_f\left( t \right) |f\in \mathcal{F} \right\} $ and $\mathbf{v}\left( t \right) =\left\{ \mathbf{v}_f\left( t \right) |f\in \mathcal{F} \right\} $.
Therefore, the optimization problem is formulated as
\begin{subequations}
	\label{eq:primalProblem}
	\begin{align}
		\begin{split}
			(\mathcal{P}) \quad
			&
			\underset{\mathbf{q},\left\{ \mathbf{\Upsilon } \left( t \right) ,t\in \mathcal{T} \right\}}
			{\min} \quad
			\frac{1}{T}\sum_{t\in \mathcal{T}}{\left( C_P\left( t \right) + \rho C_B\left( t \right) \right)} \\
			\text{s.t.} \quad
			& \eqref{eq:cachingStorage}-\eqref{eq:powerConst},
		\end{split}
		\\
		\begin{split}
			\label{const:minimumSINR}
			& \Gamma_n (t) \ge \gamma _{f},\forall n\in \pi _f(t),f\in \mathcal{F},t\in \mathcal{T},
		\end{split}
		\\
		\begin{split}
			\label{const:backhaulCap}
			& \sum_{f\in \mathcal{F}}{e_{f,i}\left( t \right) \left( 1-q_{f,i}\left( t \right) \right) R_f\left( t \right) \le}C_{i}^{BH}, \\
			& \hspace{3cm} \forall i\in \left\{ \mathcal{L} ,0 \right\} ,t\in \mathcal{T},
		\end{split}
	\end{align}
\end{subequations}
where $C_{i}^{BH}, \forall i\in \left\{ \mathcal{L},0 \right\}$ denotes the transmission capacity of the backhaul link from the cloud server to AP $i$.
Constraint \eqref{const:minimumSINR} indicates the minimum SINR requirement for a transmitted message to be successfully decoded, and constraint \eqref{const:backhaulCap} represents the transmission capacity limit for each backhaul link.

\textbf{\emph{Challenges:}}
Solving problem $\mathcal{P}$ presents several challenges.
Firstly, the lack of future information, such as content requests and user channel state information (CSI) over the next $T$ time slots, complicates cache placement updates.
Secondly, problem $\mathcal{P}$ is a mixed-integer nonlinear programming (MINLP) problem characterized by non-convex constraints, including beamforming \eqref{eq:beamformingConst} and minimum SINR requirements \eqref{const:minimumSINR}, and integer variables $\mathbf{q}$ and $\mathbf{e}(t)$.
Such MINLP problems are generally NP-hard, indicating that solving problem $\mathcal{P}$ to obtain the global optimal solution within polynomial time is extremely challenging, even with full future information \cite{floudas1995nonlinearmixed}.
In addition, problem $\mathcal{P}$ involves mixed timescale decisions; cache placement decisions made every $T$ time slots are intertwined with variables like beam direction, AP clustering, and multicast beamforming, which are optimized more frequently, further adding to its complexity.

\subsection{Two-Stage Problem Decomposition}
\label{sec:problemDecomposition}
Considering the high probability of consistent popularity of content across consecutive caching periods \cite{sun2020qoedriventransmissionaware,wu2020jointlongterm}, it is feasible to design the cache placement strategy for future requested contents of users by leveraging the historical content request information and CSI of users \cite{xiang2018cacheenabledphysical,han2022jointcache}.
Consequently, a two-stage solution framework is proposed to address the primal problem $\mathcal{P}$.
In detail, the cache placement is optimized at time $t=\theta T (\theta \in \mathbb{Z}^{+})$ in the cache placement stage.
In the content delivery stage, for a given cache placement strategy, the beam direction control, AP clustering, and multicast beamforming strategies $\mathbf{\Upsilon }\left( t \right)  =\left\{ \mathbf{c}^{SB}(t),\mathbf{e}\left( t \right),\mathbf{w}\left( t \right),\mathbf{v}\left( t \right) \right\}$ are optimized based on current CSI and content requests in each time slot.

\subsubsection{Long-Term Cache Placement Problem Formulation}
To address the lack of data on future content requests and CSI of users, a historical data-driven approach \cite{xiang2018cacheenabledphysical,sun2020qoedriventransmissionaware,wu2020jointlongterm, han2022jointcache} is employed to optimize cache placement at time $t=\theta T (\theta \in \mathbb{Z}^{+})$.
This approach uses the content requests and CSI of users from the previous $T-1$ time slots to help optimize the cache placement strategy for the subsequent $T-1$ time slots.
As a result, the cache placement problem at time $t$, e.g., $t = T$, to minimize the total network cost is formulated as follows
\begin{equation}
	\label{eq:cachePlacement}
	\begin{aligned}
		({\mathcal{P}1}) \quad
		 &
		\underset{\mathbf{q},\left\{ \mathbf{\Upsilon } \left( t \right) ,t\in \mathcal{T} ^{\prime} \right\} }
		{\min} \quad
		\frac{1}{T}\sum_{t\in \mathcal{T} ^{\prime}}{\left( C_P\left( t \right) + \rho C_B\left( t \right) \right)}
		\\
		\text{s.t.} \quad
		 & \eqref{eq:cachingStorage}-\eqref{eq:powerConst}, \eqref{const:minimumSINR}, \eqref{const:backhaulCap},
	\end{aligned}
\end{equation}
where $\mathcal{T} ^{\prime}=\left[ t-T,t-1 \right] $.

\subsubsection{Short-Term Content Delivery Problem Formulation}
Upon determining the cache placement strategy at time $\theta T (\theta \in \mathbb{Z}^{+})$, the subsequent step focuses on optimizing the content delivery strategy for each time slot $t\in \mathcal{T}$.
This optimization stage assumes that at the beginning of each time slot, the network management system is informed of the user content requests and CSI.
Then, the short-term content delivery problem is formulated as follows
\begin{equation}
	\label{eq:contentDelivery}
	\begin{aligned}
		({\mathcal{P}2}) \quad
		 &
		\underset{\mathbf{\Upsilon } \left( t \right)  }
		{\min} \quad
		C_P\left( t \right) + \rho C_B\left( t \right)
		\\
		\text{s.t.} \quad
		 & \eqref{eq:cachingStorage}-\eqref{eq:powerConst}, \eqref{const:minimumSINR}, \eqref{const:backhaulCap}.
	\end{aligned}
\end{equation}

It is evident that both ${\mathcal{P}1}$ and ${\mathcal{P}2}$ are non-convex MINLP problems, making it challenging to find their optimal solutions.

\subsection{Two-Stage Solution Framework}

\begin{figure}[htb]
	\centering
	\includegraphics[width=0.485\textwidth]{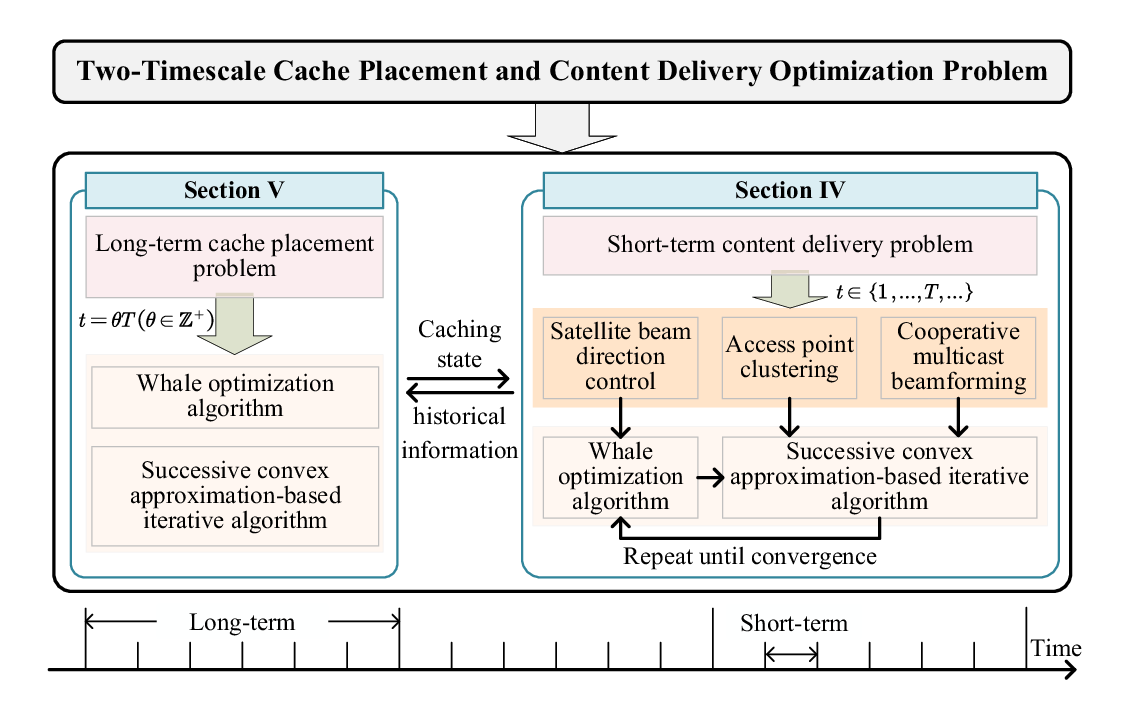}
	\caption{Overview of the proposed content caching and delivery solution.}
	\label{fig:solutionDesign}
\end{figure}

To address both the long-term cache placement problem and the short-term content delivery problems, we propose a two-stage solution framework based on different timescales, as illustrated in Fig. \ref{fig:solutionDesign}.
Briefly, after determining the cache placement state, an algorithm based on whale optimization and successive convex approximation is developed to iteratively solve the content delivery problem.
Detailed information on the solution for this stage is provided in Section \ref{sec:sol4shortTerm}.
For the cache placement problem, we design an alternating optimization algorithm that utilizes historical content requests and channel state information.
This algorithm is guided by the solution for the content delivery problem.
Details regarding the cache placement solution are presented in Section \ref{sec:sol4longTerm}.

\section{Solution Design for Short-Term Content Delivery Problem}
\label{sec:sol4shortTerm}

In this section, an alternating optimization-based solution is proposed to tackle the short-term content delivery problem for each time slot with the given cache placement strategy.
To solve problem ${\mathcal{P}2}$, we first decompose it into two subproblems, namely beam direction control and cooperative multicast beamforming.
For the beam direction control subproblem, we devise an improved whale optimization algorithm to optimize the coordinates of satellite beam centers effectively.
Subsequently, we propose a successive convex approximation-based algorithm to handle the cooperative multicast beamforming subproblem.
The relation among these subproblems is illustrated in Fig. \ref{fig:solutionDesign}.

\subsection{WOA-based Solution for Beam Direction Control}
\label{subsec:beamDirectionSubp}
With the given cache placement strategy, as well as fixed AP clustering and multicast beamforming strategies, the objective of problem ${\mathcal{P}2}$ becomes decoupled from the variable of beam direction.
Therefore, we aim to maximize the sum of user data rates in the beam direction control problem for each time slot, which is formulated as follows
\begin{equation}
	\label{eq:beamDirectionCon}
	\begin{aligned}
		 & \hspace{0.3in} ({\mathcal{P}2.1}) \quad \underset{ \mathbf{c}^{SB}(t) }{\max} \quad
		\sum_{n\in \mathcal{N}}{R_n\left( t \right) }
		\\
		 & \text{s.t.} \quad
		\Gamma _n\left( \mathbf{c}^{SB}\left( t \right) \right) \ge \gamma _f,\forall n\in \pi _f(t),f\in \mathcal{F} ,t\in \mathcal{T} ,
	\end{aligned}
\end{equation}
where $R_n\left( t \right)  =B\log \left( 1+\Gamma _n\left( \mathbf{c}^{SB}\left( t \right) \right)  \right) $ denotes the data rate of user $n$ at time slot $t$.
It is worth noting that for notation simplification in other parts of the content, $\Gamma _n\left( \mathbf{c}^{SB}\left( t \right) \right) $ is denoted as $\Gamma _n(t)$.
The above problem is a non-convex problem due to the non-convex constraint \eqref{eq:channelTransGain}, which is hard to solve by traditional optimization tools.
Hence, WOA \cite{mirjalili2016whaleoptimization} as a bionic intelligent algorithm is employed to design a low computational complexity solution for this subproblem.

Since the beam center coordinates are located on the surface of Earth, i.e., at sea level, it is sufficient to optimize only the longitude and latitude to determine the position of the beam center.
After obtaining the latitude and longitude coordinates, these can be transformed into ECEF coordinates as follows \cite{bian2024essentialsnavigation}
\begin{equation}
	\begin{aligned}
		X & = (R + h) \cos(c_{\phi}) \cos(c_{\lambda}), \\
		Y & = (R + h) \cos(c_{\phi}) \sin(c_{\lambda}), \\
		Z & = (R + h) \sin(c_{\phi}),
	\end{aligned}
\end{equation}
where $c_{\phi}$ and $c_{\lambda}$ represent the latitude and longitude in radians, respectively, $h$ denotes the altitude in meters, $R$ is the radius of Earth in meters, typically $6.371 \times 10^6$ meters, and $(X,Y,Z)$ are the corresponding ECEF coordinates.
For the coordinates of all beam centers, we have $h=0$.
Therefore, in our enhanced WOA, which includes a set of whales $\mathcal{K}=\{1,...,K\}$, we employ uniform population individual initialization, where the position of each individual is initialized as a set of latitude and longitude coordinates.
For example, the position of the $k$-th individual whale is defined as $\mathbf{\Phi }(t) \in \mathbb{R} ^{2M\times 1}$, where the $i$-th element in $\mathbf{\Phi }(t)$ is reexpressed as $\varsigma _{k,i}\left( t \right) $ with $i \in [1,...,2M]$ and $k \in [1,...,K]$.

\subsubsection{Definition of Fitness Function}
The fitness value of each whale should accurately reflect the satisfaction of the SINR requirement.
Consequently, when there exists a user that fails to meet the SINR requirement, the fitness value is assigned as infinite.
Conversely, when all users meet the SINR requirement, the fitness value is determined as the reciprocal of the sum of user data rates.
Therefore, the fitness value of the $k$-th individual with the coordinate set $\mathbf{\Phi }_{k}(t)$ in the searching space is defined as
\begin{equation}
	\label{eq:fitnessDefinition}
	F_{k}\left( t \right)  =
	\begin{cases}
		\frac{1}{\sum\limits_{n\in \mathcal{N}}{R_n\left( t \right) }},
		 & \text{if } \sum\limits_{n\in \mathcal{N}}{\mathbb{I} \left( \Gamma _n\left( \mathbf{\Phi }_k(t) \right)\ge \gamma _f \right) =N},
		\\
		+ \infty,
		 & \text {otherwise},
	\end{cases}
\end{equation}
where $\mathbb{I} \left( \cdot \right) $ is the indicator function.

\subsubsection{Shrinking Encircling Mechanism}
Once the humpback whales identify their target prey, they adopt a strategy of encircling the prey and continually adjusting their positions to approach the target.
In the context of the WOA, the current best candidate solution is regarded as the target prey or is in proximity to the optimal solution.
Hence, these encircling behaviors can be modeled as follows \cite{mirjalili2016whaleoptimization}
\begin{subequations}
	\label{eq:shrinkingEncirclingMechanism}
	\begin{align}
		\begin{split}
			\lambda
			& =| X_1\varsigma _{k,i}^{l,\ast}\left( t \right) -\varsigma _{k,i}^{l}\left( t \right) |,
		\end{split}
		\\
		\begin{split}
			\varsigma _{k,i}^{l+1}\left( t \right)
			& =\varsigma _{k,i}^{l,\ast}\left( t \right) -\lambda X_2,
		\end{split}
	\end{align}
\end{subequations}
where $X_1=2r_1$ and $X_2=2xr_2-x$ with both $r_1\sim \mu \left( 0,1 \right)$ and $r_2\sim \mu \left( 0,1 \right)$ are uniform distributions ranging from $0$ to $1$.
In addition, $x$ is gradually decreased from 2 to 0 throughout the iterations, which can be defined as $x=2\left( 1+\exp \left( l-l_{\max}/2 \right) \right) ^{-1}$ with $l_{\max}$ being the maximum number of iterations.
$l$ and $k$ denote the $l$-th iteration and $k$-th individual whale, respectively.
$\varsigma _{k,i}^{l}\left( t \right) $ and $\varsigma _{k,i}^{l+1}\left( t \right) $ are the $i$-th element of the $k$-th individual whale in the $l$-th iteration and the $(l+1)$-th iteration, respectively.
$\varsigma _{i}^{l,\ast}\left( t \right) $ is the $i$-th element of the whale with smallest fitness value in the $l$-th iteration.

\subsubsection{Spiral Exploitation}
Whales employ a hunting strategy by spitting out bubble nets in a spiral pattern while closing in on their prey \cite{mirjalili2016whaleoptimization}, which is modeled as follows
\begin{equation}
	\label{eq:spiralExploitation}
	\varsigma _{k,i}^{l+1}\left( t \right) =\varsigma _{k,i}^{l,\ast}\left( t \right) +\lambda ^{\prime}\exp \left( b\cdot r_3 \right) \cos \left( 2\pi r_3 \right),
\end{equation}
where $\lambda ^{\prime}=| \varsigma _{k,i}^{l,\ast}\left( t \right) -\varsigma _{k,i}^{l}\left( t \right) |$ indicates the distance between prey and the $k$-th individual whale, $b$ is a constant and used to define the shape of the logarithmic spiral, and $r_3\sim \mu \left( -1,1 \right) $.

It is important to note that humpback whales can encircle their prey and perform a bubble-net attack with the spiral motion simultaneously.
Similar to \cite{mirjalili2016whaleoptimization}, there is a probability of $p^{\prime}$ of selecting either the shrinking encircling mechanism or the spiral model to update the whales' position.
Thus, these behaviors can be represented as
\begin{equation}
	\label{eq:whalePoUpdate}
	\varsigma _{k,i}^{l+1}\left( t \right) =
	\begin{cases}
		\varsigma _{k,i}^{l,\ast}\left( t \right) -\lambda X_2,
		 & \text{if } p_{1} < p^{\prime},
		\\
		\varsigma _{k,i}^{l,\ast}\left( t \right) +\lambda ^{\prime}e^{b\cdot r_{3}} \cos \left( 2\pi r_3 \right) ,
		 & \text{if } p_{1}\ge p^{\prime},
		\\
	\end{cases}
\end{equation}
where $p_{1} \sim \mu \left( 0,1 \right) $ denotes the probability.

Furthermore, to prevent the WOA from converging to a local optimal solution, we incorporate the following nonlinear perturbation strategy to update the positions of whales.
\begin{equation}
	\label{eq:randomPertur}
	\varsigma _{k,i}^{l+1}\left( t \right) =\varsigma _{k,i}^{l}\left( t \right) +\frac{\sin \left( r_4 \right)}{r_4},
\end{equation}
where $r_4\sim \mu \left( -10\pi ,10\pi \right) $.
We assume that there is a probability of $p^{\prime \prime}$ to choose between the above mechanisms or the random perturbation for updating the position.

\subsubsection{Search for Prey}
In addition, whales can search randomly according to the position of each other instead of the whale with the best fitness value, and the value of $\lambda $ is employed as the indicator for such searching.
Hence, the variant of \eqref{eq:shrinkingEncirclingMechanism} is denoted as
\begin{equation}
	\label{eq:randomSelectSearch}
	\begin{aligned}
		\lambda
		 & =| X_1\varsigma _{\mathrm{rand},i}^{l}\left( t \right) -\varsigma _{k,i}^{l}\left( t \right) |,
		\\
		\varsigma _{k,i}^{l+1}\left( t \right)
		 & =\varsigma _{\mathrm{rand},i}^{l}\left( t \right) -\lambda X_2,
	\end{aligned}
\end{equation}
where $\varsigma _{\mathrm{rand},i}^{l}\left( t \right) $ denotes the $i$-th element of the randomly selected whale in the $l$-th iteration.

\begin{algorithm}[t]
	\DontPrintSemicolon
	\SetAlgoLined
	\emph{\textbf{Initialization:}} \;
	Initiate $l_{\max}$ and $b$, as well as uniformly generate the whale population $\left\{ \mathbf{\Phi }_k(t)|\forall k\in \left[ 1,...,K \right] \right\} $  \;
	Set $l = 0$ \;
	\emph{\textbf{Procedure:}} \;
	Calculate the fitness of each whale \;
	Obtain the best position of whales $\mathbf{\Phi }^{\ast}(t)$ \;
	\While{$l < l_{\max}$}{
		\For{$k \in \left[ 1,...,K \right] $}{
			Update $x$, $r_1$, $r_2$, $r_3$, $r_4$, $X_1$, $X_2$, $p_1$, $p_2$\;
			\eIf{$p_1 < p^{\prime}$}{
				\eIf{$X_2 < 1$}{
					\eIf{$p_1 \le p^{\prime \prime}$}{
						Update the position of the current whale by \eqref{eq:shrinkingEncirclingMechanism}
					}{
						Perform a nonlinear perturbation for the position of the current whale by \eqref{eq:randomPertur} \;
					}
				}{
					Randomly select a whale from the whale population to update the position of the current whale by \eqref{eq:randomSelectSearch} \;
				}
			}{
				Update the position of the current whale by \eqref{eq:spiralExploitation} \;
			}
		}
		Check if any whale goes beyond the search space and amend it \;
		Calculate the fitness of each whale \;
		Update $\mathbf{\Phi }^{\ast}(t)$ if there is a better solution \;
		$ l = l + 1$ \;
	}
	Return $\mathbf{\Phi }^{\ast}(t)$ \;
	\caption{WOA-based beam direction control algorithm}
	\label{alg:WOAalgorithm}
\end{algorithm}

\subsubsection{WOA-based Algorithm Design}
As outlined in Algorithm \ref{alg:WOAalgorithm}, the procedural execution of the proposed WOA-based beam direction control is comprehensively detailed.
Specifically, lines 9 to 21 delineate the iterative update mechanism employed for adjusting the position of each whale within the swarm; lines 24 to 26 focus on refining the optimal whale position achieved within the iteration, ensuring that the best solution vector is continuously improved upon through the execution of the algorithm.
Based on the execution flow presented in Algorithm \ref{alg:WOAalgorithm}, the computational complexity of the algorithm can be expressed as $ \mathcal{O}(l_{max}K(2M))$, where $K$ denotes the number of whale individuals, $2M$ represents the dimension of the position of each whale individual, and $l_{max}$ is the maximum number of iterations.

\subsection{SCA-based Solution for Cooperative Multicast Beamforming}
\label{subsec:SDRbeamformingSubp}

\subsubsection{Problem Analysis and Reformulation}
Given the satellite beam direction $\mathbf{c}(t)$, the AP clustering and cooperative multicast beamforming problem at time slot $t$ is expressed as
\begin{equation}
	\label{eq:cooperativeMB}
	\begin{aligned}
		({\mathcal{P}2.2}) \quad
		 &
		\underset{\mathbf{e}\left( t \right) ,\mathbf{w}\left( t \right) ,\mathbf{v}\left( t \right)  }
		{\min} \quad
		C_P\left( t \right) + \rho C_B\left( t \right)
		\\
		 & \text{s.t.} \quad
		\eqref{eq:beamformingConst},\eqref{eq:powerConst}, \eqref{const:minimumSINR}, \eqref{const:backhaulCap}.
	\end{aligned}
\end{equation}
It is worth noting that solving problem ${\mathcal{P}2.2}$ remains challenging due to the integer variables and non-convex constraints, such as \eqref{eq:beamformingConst} and \eqref{const:minimumSINR}.
To address this, we first relax the integer variables as continuous ones, e.g., treating $e_{f,m}^{\prime}(t) \in \{0,1\}$ as $0 \le e_{f,m}^{\prime}(t) \le 1$, and then proceed the following analysis and reformulation.
We define two auxiliary variables $\mathbf{i}_f\left( t \right) \in \mathbb{C} ^{\left( MK_s+LK_b \right) \times 1}$ and $\mathbf{j}_n\left( t \right) \in \mathbb{C} ^{\left( MK_s+LK_b \right) \times 1}$ to represent the beamforming matrix and channel matrix in the following forms
\begin{equation}
	\begin{aligned}
		\mathbf{i}_f( t) =  & [ \mathbf{w}_{1,f}(t)^H,...,\mathbf{w}_{M,f}(t)^H, \mathbf{v}_{1,f}( t) ^H,...,\mathbf{v}_{L,f}( t ) ^H] ^H,
		\\
		                    & \hspace{160pt} \forall f\in \mathcal{F}.
		\\
		\mathbf{j}_n( t ) = & [ \mathbf{h}_{1,n}(t)^H,...,\mathbf{h}_{M,n}(t)^H, \mathbf{u}_{1,n}( t ) ^H,...,\mathbf{u}_{L,n}( t ) ^H ] ^H,
		\\
		                    & \hspace{160pt} \forall n\in \mathcal{N} .
	\end{aligned}
\end{equation}
For the sake of convenience, we use $\mathbf{i}_{a,f}( t), \forall a \in  \mathcal{A} \triangleq [1, M+L]$ to represent the corresponding elements of $\mathbf{i}_f( t)$, such as $\mathbf{i}_{1,f}( t) = \mathbf{w}_{1,f}( t)$ and $\mathbf{i}_{M+1,f}( t) = \mathbf{v}_{1,f}( t)$.

Since $\mathbf{i}_f( t)$ introduces a new index $a \in \mathcal{A}$, the variable $\mathbf{q}_{f}(t)$ needs to be aligned with this new index to properly represent constraint \eqref{eq:beamformingConst}.
Considering the satellite has $M$ beams, the element $q_{f,0}(t)$ in $\mathbf{q}_{f}(t)$ is replicated $M$ times to represent the cache placement strategy for the satellite beams, denoted as $\mathbf{q}_{f,M}\left( t \right)$, as follows
\begin{equation}
	\mathbf{q}_{f,M}\left( t \right) = [q_{f,0}\left( t \right) ,...,q_{f,0}\left( t \right) ] \in \{0,1\}^{1\times M}.
\end{equation}
The cache placement strategy for all base stations is $\mathbf{q}_{f,L}\left( t \right) =\left[ q_{f,1}\left( t \right) ,q_{f,2}\left( t \right) ,...,q_{f,|L|}\left( t \right) \right] $.
Consequently, the revised representation of the cache placement strategy for all APs is given by $\mathbf{q}_{f}^{\prime}(t) = [\mathbf{q}_{f,M}\left( t \right), \mathbf{q}_{f,L}\left( t \right)]\in \left\{0,1\right\}^{1 \times (M+L)}$.
Then, we have $C_B^{\prime} ( t ) = \sum_{a\in \mathcal{A}}{\sum_{f\in \mathcal{F}}{e_{f,a}( t ) ( 1-{q^{\prime}}_{f,a}\left( t \right) ) R_f( t )}}$ and $C_P^{\prime}( t ) = \sum_{a\in \mathcal{A}}{\sum_{f\in \mathcal{F}}{\left\| \mathbf{i}_{a,f}( t ) \right\| _{2}^{2}}}$.
As a result, problem ${\mathcal{P}2.2}$ is reformulated as follows
\begin{subequations}
	\begin{align}
		\begin{split}
			& ({\mathcal{P}2.2.1}) \quad
			\underset{\mathbf{e}\left( t \right) ,\mathbf{i}\left( t \right)}{\min}\quad
			C_P^{\prime}\left( t \right) + \rho C_B^{\prime}\left( t \right)
		\end{split}
		\\
		\begin{split}
			\label{const:SINRreqP7}
			\text{s.t.}
			& \frac{\left| \mathbf{j}_n\left( t \right) ^H\mathbf{i}_f\left( t \right) \right|^2}{\sum\limits_{j\in \left\{ \mathcal{F} \backslash f \right\}}{\left| \mathbf{j}_n\left( t \right) ^H\mathbf{i}_j\left( t \right) \right|^2}+\sigma _{n}^{2}}\ge \gamma _f,\\
			& \hspace{1in} \forall f\in \mathcal{F} ,n\in \mathcal{N} ,
		\end{split}
		\\
		\begin{split}
			\label{const:satPowerP7}
			& \left\| \mathbf{i}_{a,f}\left( t \right) \right\| _{2}^{2}\le e_{f,a}\left( t \right) P_{a}^{\max},\forall a\in \mathcal{A} ,
		\end{split}
		\\
		\begin{split}
			& \sum_{a\in [1,M]}{\sum_{f\in \mathcal{F}}{\left\| \mathbf{i}_{a,f}\left( t \right) \right\| _{2}^{2}}}\le P_{S}^{\max},
		\end{split}
		\\
		\begin{split}
			& \sum_{f\in \mathcal{F}}{\left\| \mathbf{i}_{a,f}\left( t \right) \right\| _{2}^{2}}\le P_{a}^{\max},\forall a\in \mathcal{A},
		\end{split}
		\\
		\begin{split}
			\label{const:backhaulP7}
			& \sum_{f\in \mathcal{F}}{e_{f,a}\left( t \right) \left( 1-q^{\prime}_{f,a}\left( t \right) \right) R_f\left( t \right) \le}C_{a}^{BH}, \\
			& \hspace{1in} \forall a\in \mathcal{A},t\in \mathcal{T} ,
		\end{split}
	\end{align}
\end{subequations}
Obviously, problem ${\mathcal{P}2.2.1}$ is non-convex due to the SINR constraint \eqref{const:SINRreqP7}.

To address this problem, we introduce a slack variable $\eta _{f,n}\left( t \right)$ to replace the interference in \eqref{const:SINRreqP7}.
Hence, problem ${\mathcal{P}2.2.1}$ can be equivalently reformulated as
\begin{subequations}
	\begin{align}
		\begin{split}
			& \hspace{0.3in}({\mathcal{P}2.2.2}) \quad
			\underset{\mathbf{e}( t ) ,\mathbf{i}( t ), \boldsymbol{\eta }( t ) }{\min}
			\quad C_P^{\prime}\left( t \right) + \rho C_B^{\prime}\left( t \right)
			\\
			&\mathrm{s.t}.\quad \eqref{const:satPowerP7}-\eqref{const:backhaulP7},
		\end{split}
		\\
		\begin{split}
			\label{const:SINRreqP8}
			&\frac{\left| \mathbf{j}_n\left( t \right) ^H\mathbf{i}_f\left( t \right) \right|^2}{\eta _{f,n}\left( t \right)}\ge \gamma _f,\forall f\in \mathcal{F} ,n\in \mathcal{N} ,
		\end{split}
		\\
		\begin{split}
			&\sum_{j\in \left\{ \mathcal{F} \backslash f \right\}}{\left| \mathbf{j}_n\left( t \right) ^H\mathbf{i}_j\left( t \right) \right|^2}+\sigma _{n}^{2}\le \eta _{f,n}\left( t \right) , \\
			& \hspace{94pt} \forall f\in \mathcal{F} ,n\in \mathcal{N},
		\end{split}
	\end{align}
\end{subequations}
where $\boldsymbol{\eta }\left( t \right) =\left\{ \eta _{f,n}\left( t \right) |\forall f\in \mathcal{F} ,n\in \mathcal{N} \right\} $.
This problem is convex except for constraint \eqref{const:SINRreqP8}.
Subsequently, the SCA method is employed.
This method iteratively approximates the non-convex components using first-order Taylor expansions, with the aim of approaching the optimal solution.
By following this approach, the left side of \eqref{const:SINRreqP8} can be approximated as
\begin{equation}
	\label{eq:SINRTaylor}
	\begin{aligned}
		 & \frac{\left| \mathbf{j}_n\left( t \right) ^H\mathbf{i}_f\left( t \right) \right|^2}{\eta _{f,n}\left( t \right)}
		\ge \psi \left( \mathbf{i}_f\left( t \right) ,\eta _{f,n}\left( t \right) ;\mathbf{i}_{f}^{\left( l \right)}\left( t \right) ,\eta _{f,n}^{\left( l \right)}\left( t \right) \right)
		\\
		 & \triangleq \frac{2\mathrm{Re}\left( \mathbf{i}_{f}^{\left( l \right)}\left( t \right) ^H\mathbf{j}_n\left( t \right) \mathbf{j}_n\left( t \right) ^H\mathbf{i}_f\left( t \right) \right)}{\eta _{f,n}^{\left( l \right)}\left( t \right)}
		\\
		 & -\eta _{f,n}\left( t \right) \frac{\left| \mathbf{j}_n\left( t \right) ^H\mathbf{i}_{f}^{\left( l \right)}\left( t \right) \right|^2}{\eta _{f,n}^{\left( l \right)}\left( t \right) ^2}, \forall f\in \mathcal{F} ,n\in \mathcal{N},
	\end{aligned}
\end{equation}
where $\mathbf{i}_{f}^{\left( l \right)}\left( t \right) $ and $\eta _{f,n}^{\left( l \right)}\left( t \right) $ represent the values of $\mathbf{i}_{f}\left( t \right) $ and $\eta _{f,n}\left( t \right) $ obtained from the previous iteration $l$, respectively.
The right side of \eqref{eq:SINRTaylor}, which represents the first-order Taylor expansion of the left side, serves as the lower bound for the left side, and $\mathrm{Re}\left( x \right) $ denotes the real part of complex value $x$.
Then, constraint \eqref{const:SINRreqP8} can be approximated as the following convex constraint
\begin{equation}
	\label{eq:convexSINR}
	\begin{aligned}
		 & \psi \left( \mathbf{i}_f\left( t \right) ,\eta _{f,n}\left( t \right) ;\mathbf{i}_{f}^{\left( l \right)}\left( t \right) ,\eta _{f,n}^{\left( l \right)}\left( t \right) \right) \ge \gamma _f,
		\\
		 & \hspace{2.1in} \forall f\in \mathcal{F} ,n\in \mathcal{N} .
	\end{aligned}
\end{equation}

By applying the SCA method and substituting the non-convex constraint \eqref{const:SINRreqP8} with \eqref{eq:convexSINR} in problem ${\mathcal{P}2.2.2}$, the approximated convex optimization problem at iteration $l+1$ is given as
\begin{subequations}
	\begin{align}
		\begin{split}
			({\mathcal{P}2.2.3}) \,\,
			&
			\underset{\mathbf{e}( t ) ,\mathbf{i}( t ), \boldsymbol{\eta }( t ) }
			{\min} \,\,
			C_P^{\prime}\left( t \right) + \rho C_B^{\prime}\left( t \right)
			\\
			\mathrm{s.t}. \quad
			& \eqref{const:satPowerP7}-\eqref{const:backhaulP7},\eqref{eq:convexSINR}.
		\end{split}
	\end{align}
\end{subequations}

\subsubsection{Problem Feasibility}
\label{subsec:P22feasibility}

The feasibility of problem ${\mathcal{P}2.2}$, influenced by user channel conditions and high SINR demands, is not guaranteed.
However, assessing the feasibility of the NP-hard problem ${\mathcal{P}2.2}$ poses a significant challenge, comparable in complexity to solving the problem itself \cite{tao2016contentcentricsparse}.
In the related studies \cite{xiang2013coordinatedmulticast,tao2016contentcentricsparse}, the feasibility of SINR-based multi-cell multicast beamforming optimization problem is explored and discussed, and the essential conditions for their feasibility have been identified.
In this paper, for the sake of simplicity, we focus on a feasible case of problem $\mathcal{P}$.

\subsubsection{SCA-based Algorithm Design}

\begin{algorithm}[t]
	\DontPrintSemicolon
	\SetAlgoLined
	\emph{\textbf{Initialization:}} \;
	Set $l=0$ \;
	Find the feasible starting points $\left\{ \mathbf{i}\left( t \right) ^{\left( l \right)},\boldsymbol{\eta }\left( t \right) ^{\left( l \right)} \right\}$ \;
	\emph{\textbf{Procedure:}} \;
	\Repeat{Convergence}{
		Solve ${\mathcal{P}2.2.3}$ using Gurobi to obtain the solution $\left\{ \mathbf{e}\left( t \right) ^*,\mathbf{i}\left( t \right) ^*,\boldsymbol{\eta }\left( t \right) ^* \right\} $ \;
		$l=l+1$ \;
		Update $\mathbf{i}\left( t \right) ^{\left( l \right)}=\mathbf{i}\left( t \right) ^*$, and $\boldsymbol{\eta }\left( t \right) ^{\left( l \right)}=\boldsymbol{\eta }\left( t \right) ^*$ \;
	}
	\caption{SCA-based cooperative multicast beamforming algorithm}
	\label{alg:SCAalgorithm}
\end{algorithm}

Relaxing the integer variables to continuous variables transforms problem ${\mathcal{P}2.2.3}$ into a relaxed convex quadratic programming problem.
The computational complexity of solving this problem using the interior point method, which is readily available in GUROBI \cite{gurobi2024gurobioptimizer} and CPLEX \cite{ibm2024cplex}, is $\mathcal{O}(l_{max}(F(M(K_s+1)+L(K_b+1)+N))^3)$, where $l_{max}$ denotes the number of iterations required for convergence.
In its original form, problem ${\mathcal{P}2.2.3}$ with the integer variable $\mathbf{e}(t)$ is a mixed-integer quadratic constrained programming (MIQCP) problem.
In this study, GUROBI is utilized to handle problem ${\mathcal{P}2.2.3}$.
The solution obtained from solving problem ${\mathcal{P}2.2.3}$ provides an upper bound for problem ${\mathcal{P}2.2.1}$.
The algorithm designed to solve the joint multicast beamforming problem ${\mathcal{P}2.2.3}$ is detailed in Algorithm \ref{alg:SCAalgorithm}.

\section{Solution Design for Long-Term Cache Placement Problem}
\label{sec:sol4longTerm}

It is evident that problem ${\mathcal{P}1}$ is a non-convex MINLP problem, which is difficult to solve directly.
Based on the transformation from the above subsection, the long-term cache placement problem can be reformulated in a similar way to problem ${\mathcal{P}2}$ as
\begin{subequations}
	\label{eq:longTermCP}
	\begin{align}
		\begin{split}
			& ({\mathcal{P}1.1}) \quad
			\underset{{\mathbf{q}^{\prime}}, \mathbf{e},\mathbf{i},\mathbf{c}}
			{\min}\,\,
			\frac{1}{T}\sum_{t \in \mathcal{T}}{\left( C_P^{\prime}\left( t \right) + \rho C_B^{\prime}\left( t \right) \right)}
		\end{split}
		\\
		\begin{split}
			\label{const:SINRreqP11}
			\text{s.t.} \quad
			& \frac{\left| \mathbf{j}_n\left( t \right) ^H\mathbf{i}_f\left( t \right) \right|^2}{\sum\limits_{j\in \left\{ \mathcal{F} \backslash f \right\}}{\left| \mathbf{j}_n\left( t \right) ^H\mathbf{i}_j\left( t \right) \right|^2}+\sigma _{n}^{2}}\ge \gamma _f, \\
			& \hspace{77pt} \forall f\in \mathcal{F} ,n\in \mathcal{N} , t \in \mathcal{T},
		\end{split}
		\\
		\begin{split}
			\label{const:coupledPowerClusteringP11}
			& \left\| \mathbf{i}_{a,f}\left( t \right) \right\| _{2}^{2}\le e_{f,a}\left( t \right) P_{a}^{\max},\forall a\in \mathcal{A}, \forall t \in \mathcal{T},
		\end{split}
		\\
		\begin{split}
			\label{const:satellitePowerP11}
			& \sum_{a\in [1,M]}{\sum_{f\in \mathcal{F}}{\left\| \mathbf{i}_{a,f}\left( t \right) \right\| _{2}^{2}}}\le P_{S}^{\max}, t \in \mathcal{T},
		\end{split}
		\\
		\begin{split}
			\label{const:APpowerP11}
			& \sum_{f\in \mathcal{F}}{\left\| \mathbf{i}_{a,f}\left( t \right) \right\| _{2}^{2}}\le P_{a}^{\max},\forall a\in \mathcal{A}, t \in \mathcal{T},
		\end{split}
		\\
		\begin{split}
			\label{const:backConstP11}
			& \sum_{f\in \mathcal{F}}{e_{f,a}\left( t \right) \left( 1-q^{\prime}_{f,a}\left( t \right) \right) R_f\left( t \right) \le}C_{a}^{BH}, \\
			& \hspace{77pt} \forall a \in \mathcal{A} ,t\in \mathcal{T} ,
		\end{split}
		\\
		\begin{split}
			\label{const:cacheStorageP11}
			& \sum_{f\in \mathcal{F}}{q^{\prime}_{f,a}\left( t \right)}\le \phi _a,\forall a\in \mathcal{A} ,t\in \mathcal{T} ,
		\end{split}
	\end{align}
\end{subequations}
where $ \mathbf{q}^{\prime} = \{ \mathbf{q}_{f}^{\prime}(t) \in \left\{0,1\right\}^{1 \times (M+L)} | f \in \mathcal{F}, t \in \mathcal{T}\}$, which is consistent with the definition in problem $\mathcal{P}2.2.1$.
It is noted that $\mathbf{q}^{\prime}$ is updated every $T$ time slots, e.g., $t\in \left\{ \theta T|\theta \in \mathbb{Z} + \right\} $.

Similar to problem ${\mathcal{P}2.2.1}$, we relax the integer variables as continuous ones, e.g., relaxing $q_{f,a}^{\prime}(t) \in \{0,1\}$ as $0 \le q_{f,a}^{\prime}(t) \le 1$.
We then employ an alternating method to address the relaxed problem.
Initially, the beam direction strategy can be obtained using Algorithm \ref{alg:WOAalgorithm}.
After that, the following problem is expressed as
\begin{subequations}
	\label{eq:longTermCPnoBDC}
	\begin{align}
		({\mathcal{P}1.2}) \quad
		                  &
		\underset{{\mathbf{q}^{\prime}}, \mathbf{e},\mathbf{i}}
		{\min}\,\,
		\frac{1}{T}\sum_{t \in \mathcal{T}}{(C_P^{\prime}\left( t \right) + \rho C_B^{\prime}\left( t \right))}
		\\
		\text{s.t.} \quad & \eqref{const:SINRreqP11}-\eqref{const:cacheStorageP11}.
	\end{align}
\end{subequations}
Then, the non-convex SINR constraint \eqref{const:SINRreqP11} is transformed into a convex one by employing a first-order Taylor expansion, as detailed in \eqref{eq:SINRTaylor}.
In addition, considering the coupled variables $\mathbf{e}$ and ${\mathbf{q}}^{\prime}$ in objective function and \eqref{const:backConstP11}, we fix the variable $\mathbf{e}$ in iteration $(l+1)$ to its determined values $\mathbf{e}^{(l)}$ obtained from iteration $l$.
This adjustment renders the objective function and \eqref{const:backConstP11} convex in each iteration.
As a result, the approximated convex optimization problem for ${\mathcal{P}1.2}$ at iteration $(l+1)$ can be expressed as
\begin{subequations}
	\label{eq:SCAlongTermCP}
	\begin{align}
		\begin{split}
			({\mathcal{P}1.3}) \quad
			&
			\underset{{\mathbf{q}}^{\prime}, \mathbf{e},\mathbf{i},\boldsymbol{\eta}}
			{\min}\,\,
			\frac{1}{T}\sum_{t \in \mathcal{T}}{(C_P^{\prime}\left( t \right) + \rho C_B^{\prime}\left( t \right))}
		\end{split}
		\\
		\begin{split}
			\text{s.t.} \quad
			& \psi \left( \mathbf{i}_f\left( t \right) ,\eta _{f,n}\left( t \right) ;\mathbf{i}_{f}^{\left( l \right)}\left( t \right) ,\eta _{f,n}^{\left( l \right)}\left( t \right) \right) \ge \gamma _f, \\
			& \hspace{1in} \forall f\in \mathcal{F} ,n\in \mathcal{N} ,
		\end{split}
		\\
		\begin{split}
			& \sum_{f\in \mathcal{F}}{e^{(l)}_{f,a}\left( t \right) \left( 1-q^{\prime}_{f,a}\left( t \right) \right) R_f\left( t \right) \le}C_{a}^{BH}, \\
			& \hspace{1in} \forall a\in \mathcal{A} ,t\in \mathcal{T} , \\
			& \eqref{const:coupledPowerClusteringP11}, \eqref{const:satellitePowerP11}, \eqref{const:APpowerP11}, \eqref{const:cacheStorageP11},
		\end{split}
	\end{align}
\end{subequations}
where $\mathbf{i}_{f}^{\left( l \right)}\left( t \right)$, $\eta _{f,n}^{\left( l \right)}\left( t \right)$ and $e^{(l)}_{f,i}$ are the value obtained from iteration $l$.

By relaxing the integer variables to continuous ones, problem $\mathcal{P}1.3$ is obviously a relaxed convex quadratic programming problem, with a computational complexity of $\mathcal{O}(l_{max}((TF(M(K_s+1)+L(K_b+1)+N)+F(M+L))^3))$ using interior point method, where $l_{max}$ is the number of iterations required for convergence.
Furthermore, it is evident that problem ${\mathcal{P}1.3}$ without the relaxation of the integer variables is an MIQCP, identical to problem $\mathcal{P}2.2.3$.
Similarly, GUROBI is used to handle this problem.
During the alternating optimization process, the bound of problem ${\mathcal{P}1.3}$ is progressively tightened to approach the optimal value of problem ${\mathcal{P}1.2}$ until convergence is achieved.

\begin{table}[ht]
	\caption{Main Simulation Parameters}
	\centering
	\label{tab:simulationSettings}
	\begin{tabular}{c|c}
		\toprule[1pt]
		\textbf{Parameter}                                      & \textbf{Value}         \\ \midrule[1pt]
		The number of users, $N$                                & 12                     \\
		The number of satellite beams, $Q$                      & 2                      \\
		The number of base stations, $B$                        & 4                      \\
		The number of satellites, $M$                           & 1                      \\
		The number of contents, $F$                             & $10$                   \\
		Carrier Frequency,$f_{c}$                               & $4\,\rm{GHz}$ (C band) \\
		Beam bandwidth, $B$                                     & $10\,\rm{MHz}$         \\
		The number of antennas of BS, $K_b$                     & 2                      \\
		The number of feeds of satellite, $K_s$                 & 8                      \\
		The altitude of satellite orbits, $H_{q}$               & $1000\,km$             \\
		The number of initial whales, $K$                       & $81$                   \\
		The shape of the spiral, $b$                            & $1$                    \\
		Probability of spiral exploitation, $1-p^{\prime}$      & $0.5$                  \\
		Probability of shrinking encircling, $p^{\prime\prime}$ & $0.9$                  \\
		The power spectral density of noise                     & $-174\,\rm{dBm/Hz}$    \\
		The transmission power of satellite, $P_s^{max}$        & $46\,\rm{dBm}$         \\
		The transmission power of BSs, $P_B^{max}$              & $43\,\rm{dBm}$         \\
		Minimum transmission rate, $R_{f}(t)$                   & $4\,\rm{Mbps}$         \\
		Configuration period, $T$                               & $10$                   \\
		Length of time slot, $\tau$                             & $1\,s$                 \\
		\bottomrule[1pt]
	\end{tabular}
\end{table}

\section{Simulation Results}
\label{sec:simulation}

In this section, we outline the simulation scenario and the settings for the parameters.
Subsequently, we present the results of these simulations, which aim to evaluate the network costs in dynamic cache-aided satellite-terrestrial networks employing our proposed two-stage strategies for joint cache placement, beam direction control, and multicast beamforming.
The AGI Systems Tool Kit creates a dynamic satellite-terrestrial network topology, which is detailed in Table \ref{tab:simulationSettings} alongside the key simulation parameters.

\begin{figure}[tb]
	\centering
	\includegraphics[width=0.48\textwidth]{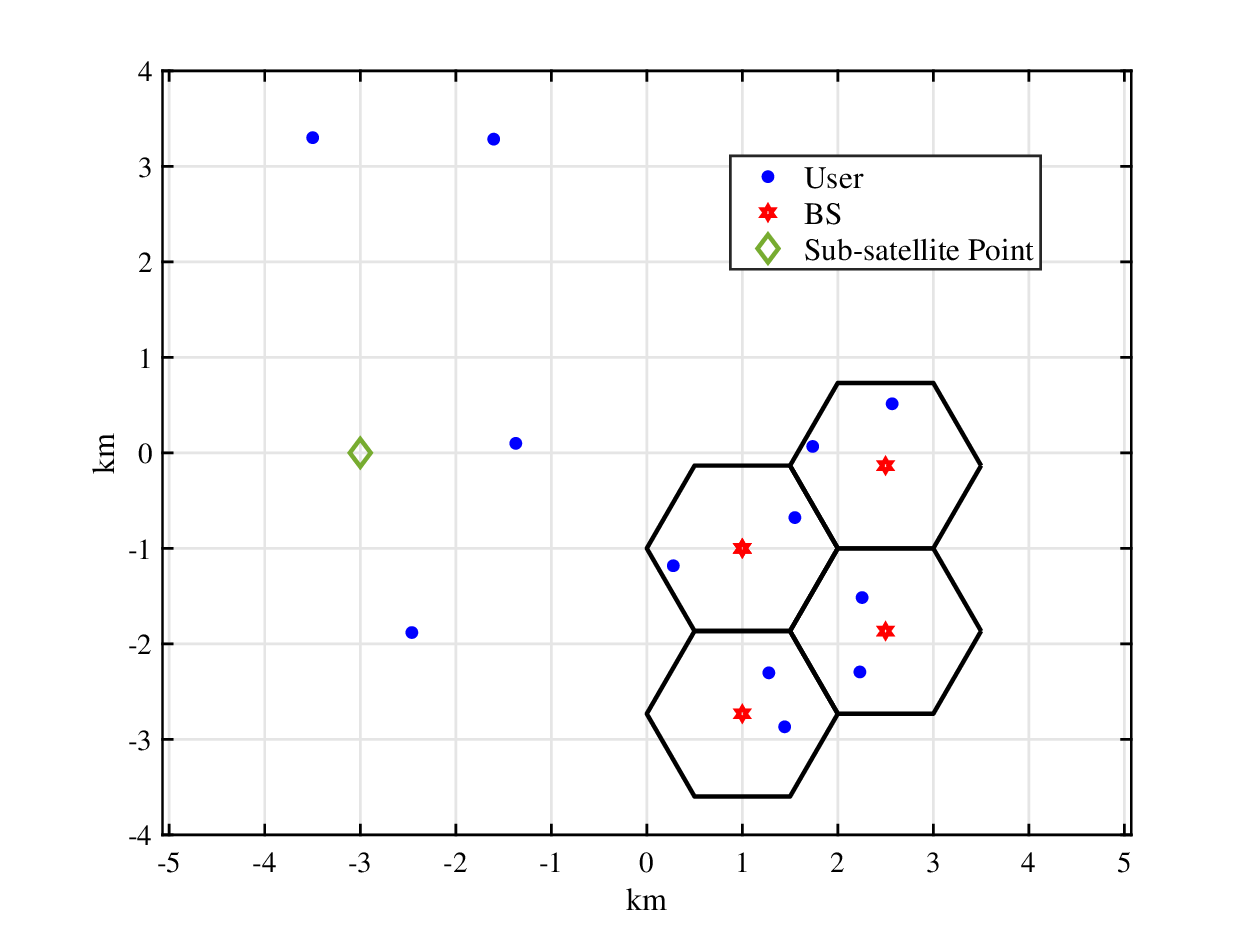}
	\caption{The network topology at the fifth time slot. Dots represent users, hexagons denote base stations, and a diamond signifies the sub-satellite point.}
	\label{fig:des_simNetTopology}
\end{figure}

Our simulations utilize a Walker satellite constellation with 20 orbits, each at a 1000-kilometer altitude and a $45^\circ$ inclination, hosting 30 satellites per orbit.
The targeted service area, centered at (41.7642$^\circ$ N, 86.6513$^\circ$ E), includes 4 base stations and 12 users, covered by at least one satellite.
Considering the high velocity of satellites at an altitude of 1000 $km$, the duration of each time slot is set to one second to ensure the feasibility of the quasi-static setting in the time slot-based topology modeling.
In addition, to simulate the dynamic characteristics of STNs, the cache placement strategy is updated every ten seconds.
The network topology for the fifth time slot is depicted in Fig. \ref{fig:des_simNetTopology}.
Satellites have a maximum transmission power of 46 dBm with 8 antennas each, while base stations have 43 dBm and 4 antennas each.
The path loss for base station communication channels is given by $PL(dB)=148.1 + 37.6 \log_{10}(d)$, where $d$ is the distance in kilometers between a base station and a user.
Shadow fading is modeled with a log-normal distribution at 8dB, and base station antenna gain is 10 dBi \cite{han2022jointcache,tao2016contentcentricsparse,wu2020jointlongterm}.
Noise power spectral density is set at -174 dBm/Hz for all users.
Both satellites and base stations operate in the 4GHz frequency band, using a 10MHz bandwidth.
Successful content delivery necessitates a minimum transmission rate of 4Mbps.
Unless noted otherwise, the content library contains 10 items, with each base station having a cache capacity of 10\% of the total content library size; satellites with two beams, which are considered to have two access points, offer a cache capacity of 20\% of the total content library size; and the trade-off factor between backhaul traffic and transmission power consumption is set to 1.
Simulations assume a non-uniform distribution of popularity: 30\% of items, considered more popular, have a collective request probability of 0.5, while the remaining 70\% also have a total request probability of 0.5, following a Zipf distribution with a shape parameter of 1.

To validate the effectiveness of the proposed algorithms, the following three baselines are considered for a comparative analysis of simulation results:
\begin{enumerate}
	\item \emph{Popularity-Aware Caching (PAC):}
	      With a known distribution of content popularity, each base station and satellite leverages its full caching capacity to store the most popular content.
	      This results in the same contents across all base stations, attributed to equal cache sizes.
	\item \emph{Fixed Beam Direction (FBD):}
	      The direction of satellite beams is fixed to the satellite position, incorporating the cache placement and multicast strategy as designed in this paper.
	\item \emph{Unicast Content Delivery (UCD):} In the absence of multicast transmission, every base station and satellite engages in unicast delivery of content to users \cite{yuan2024jointbeam}, employing a probabilistic caching strategy \cite{tao2016contentcentricsparse} and beam direction control as designed in this paper.
\end{enumerate}

\begin{figure}[t]
	\centering
	\includegraphics[width=0.45\textwidth]{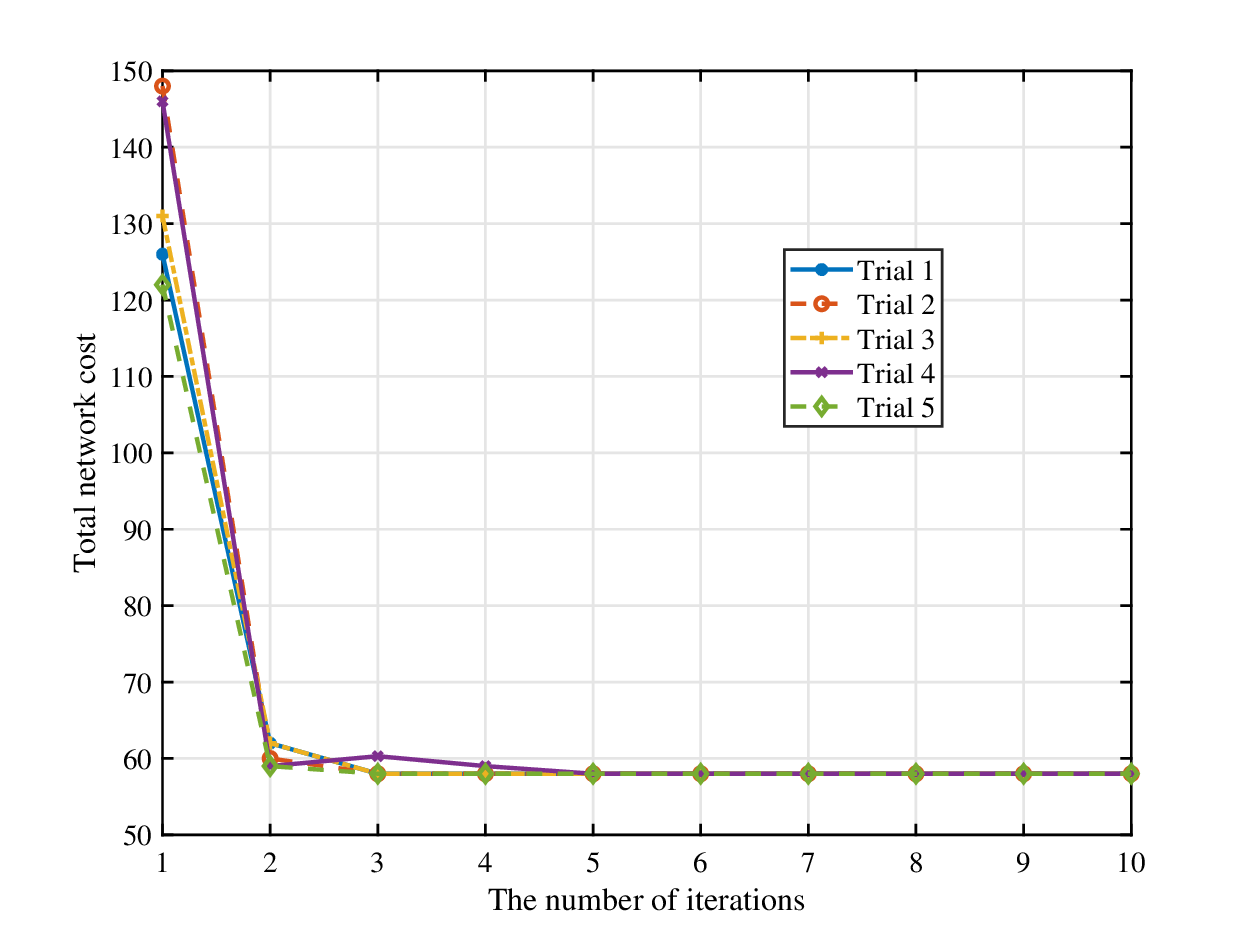}
	\caption{Convergence behavior of our proposal.}
	\label{fig:eva_convergence}
\end{figure}

To validate the convergence efficiency of our proposed alternating optimization-based content delivery algorithm in satellite-terrestrial networks, Fig. \ref{fig:eva_convergence} presents results from five different trials, each with distinct initial setups.
Each trial begins by initializing the multicast beamforming variables $\mathbf{w}(t)$ and $\mathbf{v}(t)$ using a complex Gaussian distribution, while ensuring the constraint \eqref{eq:powerConst} is satisfied.
In addition, the elements of the access point clustering matrix $\mathbf{e}(t)$ are randomly assigned 0 or 1.
As shown in Fig. \ref{fig:eva_convergence}, despite diverse initial conditions for multicast clustering and beamforming, the optimization objective values demonstrate a generally decreasing trend and stabilize at a consistent level after five iterations.
Initial fluctuations in objective values across trials are noted, attributed to the interaction between beam direction control and multicast clustering, thus affecting cooperative multicast beamforming strategies and leading to performance variances.
After several iterations, with the stable beam direction, the optimization objective value begins to decrease, and by the fifth iteration, it clearly showed that the objective values of all the trials converged to a stable value, thus demonstrating the convergence efficiency of our proposed algorithms.

\begin{figure}[t]
	\centering
	\includegraphics[width=0.45\textwidth]{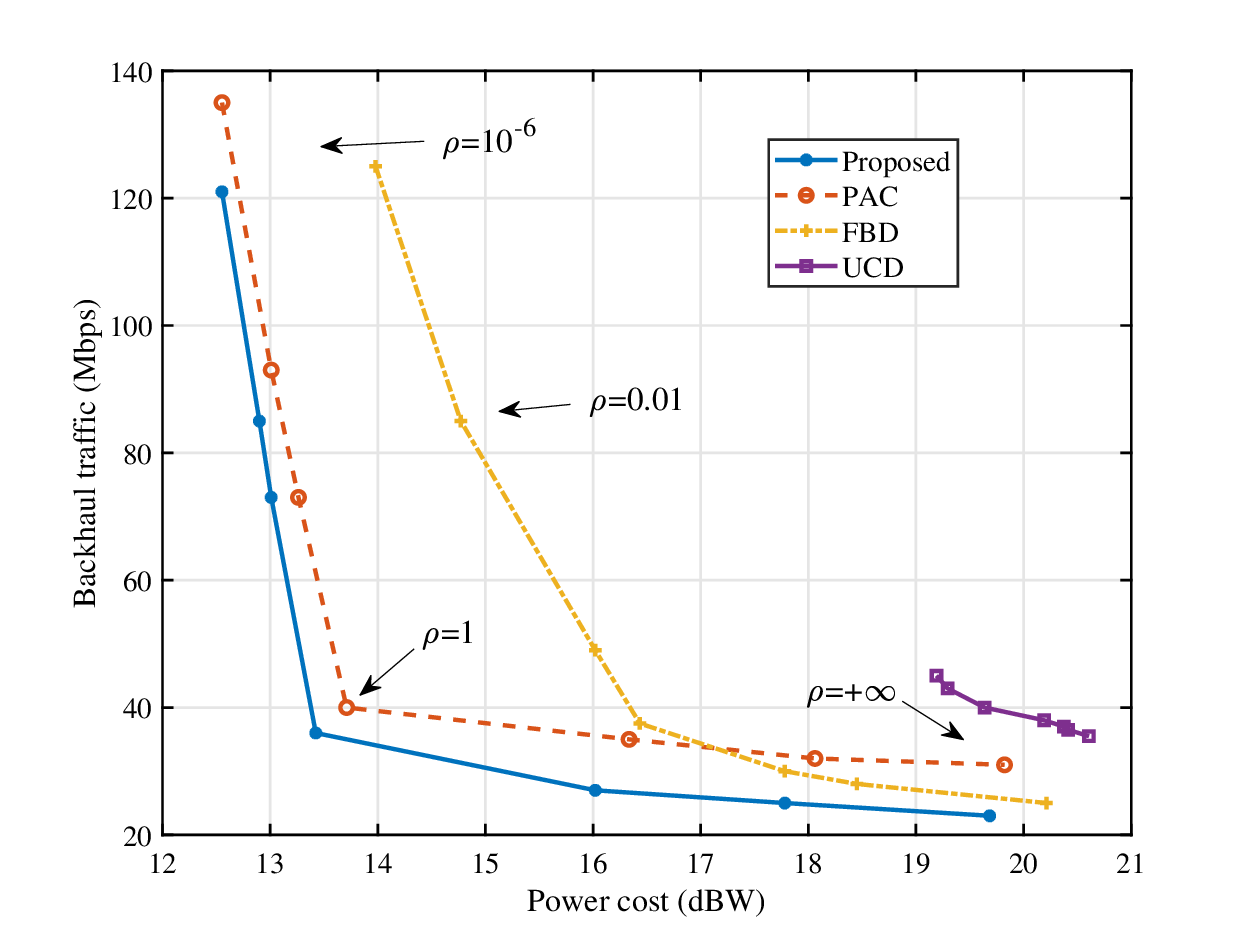}
	\caption{Trade-offs between power cost and backhaul traffic with different $\rho$.}
	\label{fig:eva_balance}
\end{figure}

\begin{figure}[t]
	\centering
	\includegraphics[width=0.45\textwidth]{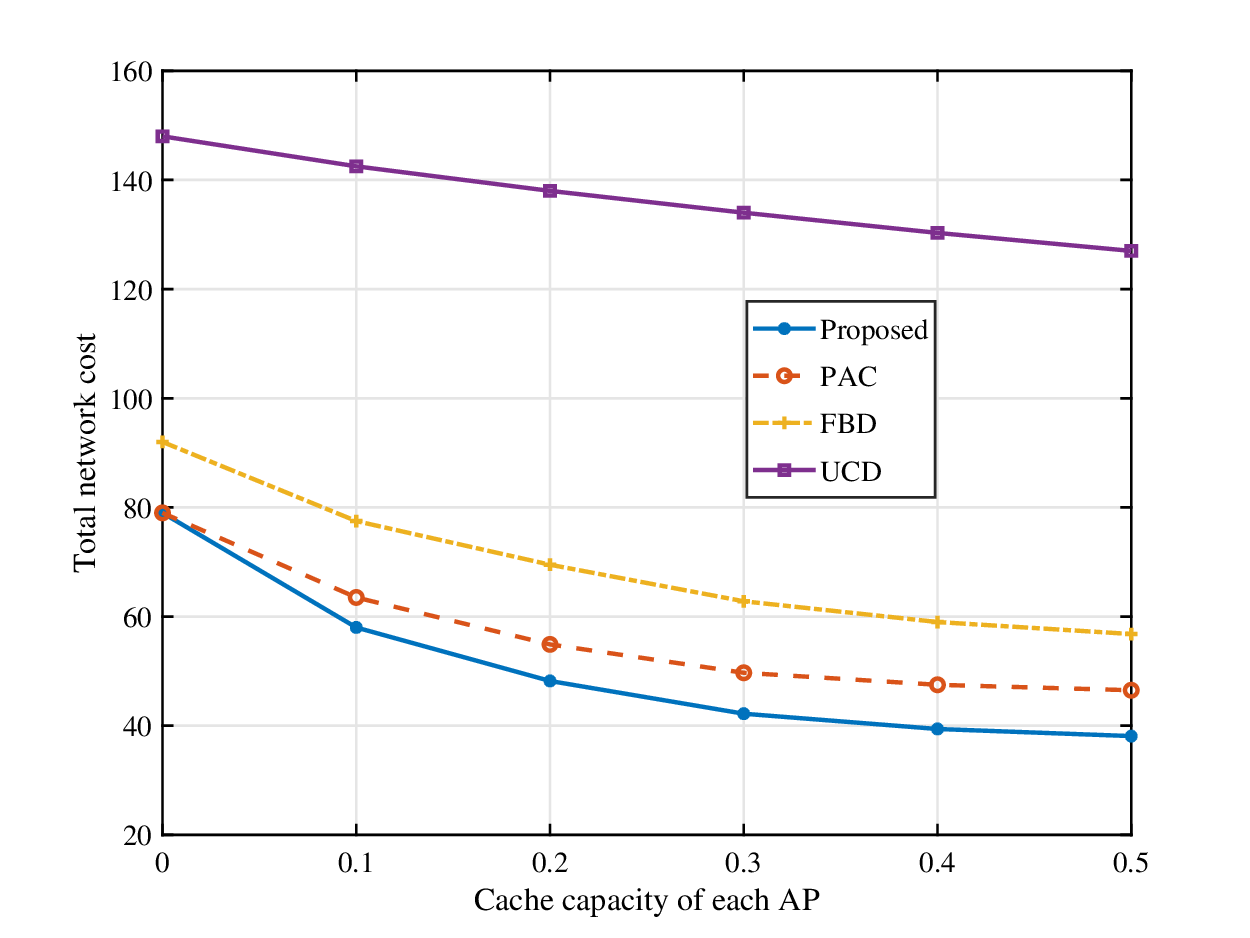}
	\caption{The impact of cache capacity at each AP on the total network cost.}
	\label{fig:eva_objv}
\end{figure}

Fig. \ref{fig:eva_balance} illustrates the trade-off between transmission power consumption and backhaul traffic for various weighting factors, denoted by $\rho \in \{ 10^{-6},0.01,0.05,1,10,1000,+\infty \}$.
A lower value of $\rho$ indicates a diminished influence of backhaul traffic on the optimization objective, with $\rho = 10^{-6}$ favoring content transmission power optimization, and $\rho \rightarrow +\infty$ prioritizing backhaul traffic optimization.
The caching capacity is set at 10\% of the total content library size for each base station and at 20\% for each satellite.
Fig. \ref{fig:eva_balance} reveals that at lower $\rho$ values, such as $10^{-6}$, our proposal optimizes for enhanced content delivery through satellite-terrestrial cooperative multicast beamforming, thus minimizing transmission power consumption.
Conversely, as $\rho \rightarrow +\infty$, the focus shifts to optimizing backhaul traffic, which increases power consumption to meet the user rate demand.
Notably, under various weighting factor settings, the trade-off efficiency between transmission power consumption and backhaul traffic achieved by our proposed algorithm outperforms that of the three comparison schemes.
Specifically, our approach significantly outperforms the popularity-aware caching scheme, achieving the lowest content transmission power consumption and reducing backhaul traffic by approximately 13\%, demonstrating the effectiveness of our cache placement optimization strategy.
Furthermore, our proposed algorithm reduces transmission power consumption by nearly 1.5 dB compared to the fixed beam direction scheme, underscoring the critical role of satellite beam direction control in optimizing satellite-terrestrial resource allocation.

\begin{figure}[t]
	\centering
	\includegraphics[width=0.45\textwidth]{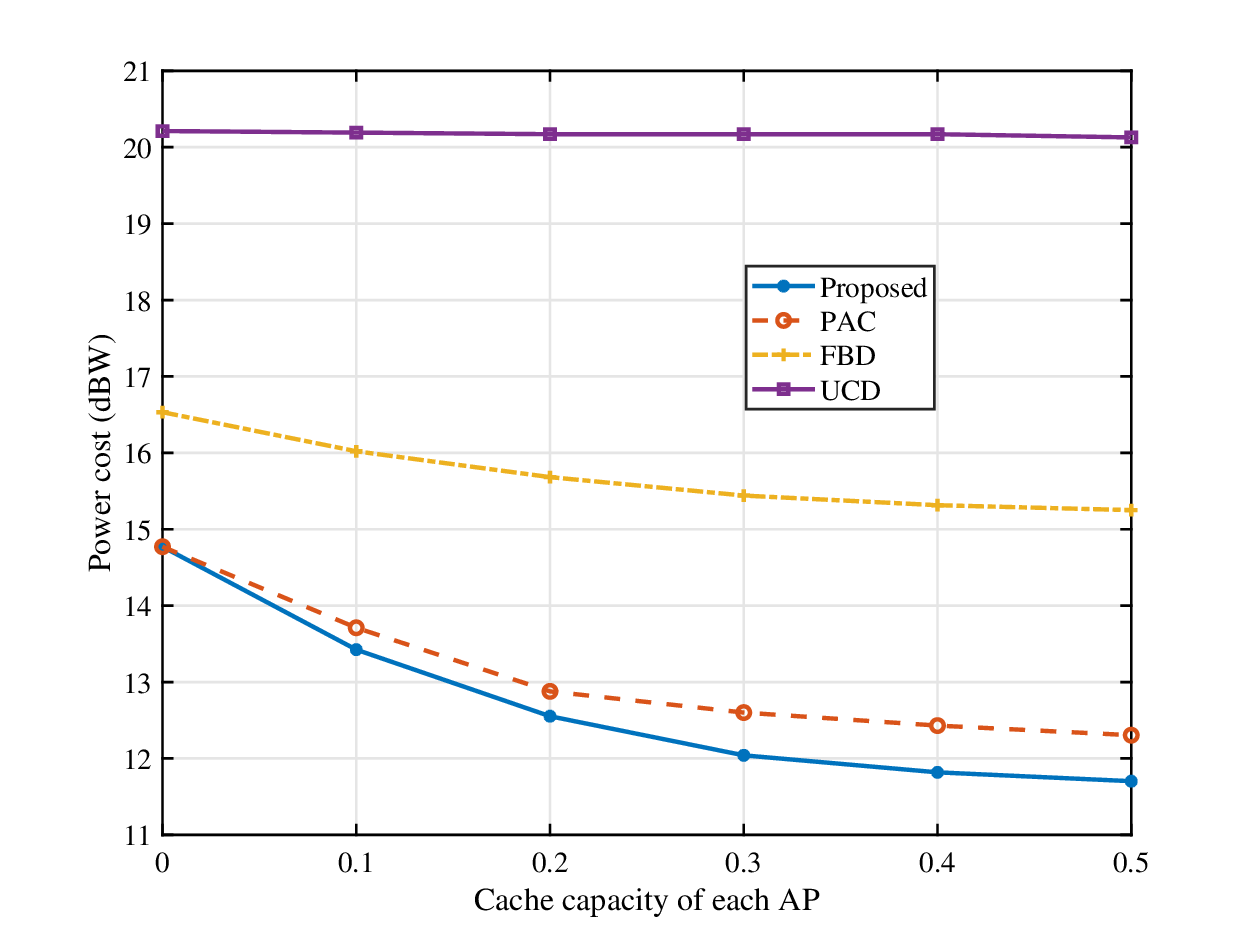}
	\caption{The impact of cache capacity at each AP on the power cost.}
	\label{fig:eva_power}
\end{figure}

\begin{figure}[t]
	\centering
	\includegraphics[width=0.45\textwidth]{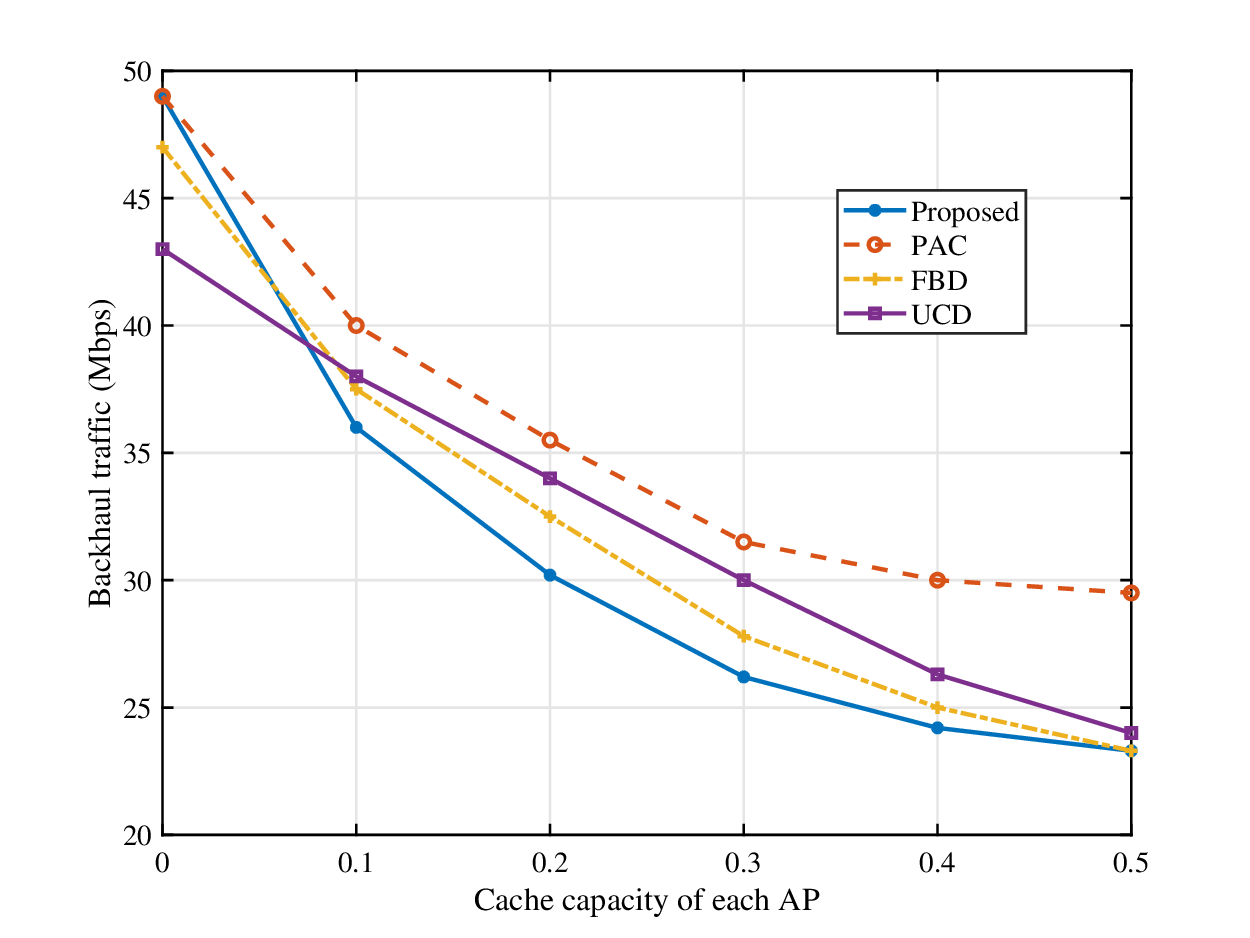}
	\caption{The impact of cache capacity at each AP on the backhaul traffic.}
	\label{fig:eva_backhaul}
\end{figure}

Fig. \ref{fig:eva_objv}, Fig. \ref{fig:eva_power}, and Fig. \ref{fig:eva_backhaul} demonstrate how varying caching capacities at network nodes, such as satellite and base stations, affect system performance.
Our simulation primarily examines scenarios with limited edge caching capacity, reflective of real-world conditions where edge nodes have considerably less storage compared to the entire cloud content library.
Specifically, we explore caching capacities at each node being $\{0, 0.1, 0.2, 0.3, 0.4, 0.5\}$ of the total content library size.
When edge nodes lack caching storage, our proposed algorithm results in an increase in backhaul traffic costs by approximately 14\% and 4\% compared to the UCD and FBD schemes, respectively.
However, as depicted in Fig. \ref{fig:eva_power}, this increase in cost is compensated by a significant reduction in transmission power consumption, with performance improvements of nearly 27\% and 12\% relative to the UCD and FBD schemes, respectively.
The enhanced power efficiency achieved by our proposed algorithm justifies the increase in backhaul traffic cost, thereby aligning with our optimization objectives.
It is evident that, when network nodes with cached content, compared to the scenario without caching, our proposed algorithm doubles the optimization objective value, reduces power consumption by 3 dB, and decreases backhaul traffic by nearly 52\%, significantly outperforming the other three schemes.
As the caching capacity of the nodes increases, both the transmission power consumption and backhaul traffic decrease accordingly.
Moreover, the incremental performance gains from increased caching capacity begin to level off, suggesting a diminishing necessity for large edge caching capacities due to the effectiveness of the cache placement optimization strategy.
Notably, with caching 50\% of the content at edge nodes, our proposal achieves a reduction of 3.2 dB in power consumption compared to the fixed beam direction, and a nearly 22\% decrease in backhaul traffic compared to popularity-aware caching.
These highlight the critical role and efficiency of beam direction control and cache placement optimization strategies.

\section{Conclusions}
\label{sec:conclusions}

In this paper, we have focused on an edge caching and content delivery scenario within the multi-beam satellite-terrestrial network and have formulated a two-timescale problem for joint cache placement and cooperative multicast beamforming optimization.
The main objective is to minimize the network cost in terms of transmission power consumption and backhaul traffic.
To tackle this problem with mixed timescales, we have decoupled it into a short-term content delivery problem, including beam direction, access point clustering, and multicast beamforming, as well as into a long-term cache placement problem.
Then, an alternating optimization algorithm inspired by whale optimization and successive convex approximation methods has been designed to solve the former, while an iterative algorithm leveraging historical information has been proposed to address the latter.
Simulations have demonstrated the effectiveness of our proposed algorithms, showcasing their convergence and significantly reducing transmission power costs and backhaul traffic by up to 52\%.

\ifCLASSOPTIONcaptionsoff
	\newpage
\fi

\bibliographystyle{IEEEtran}
\bibliography{bibUsedinPaper,bstControlForIEEEtran}

\vspace{-0.6in}
\begin{IEEEbiography}[{\includegraphics[width=1in,height=1.25in,clip,keepaspectratio]{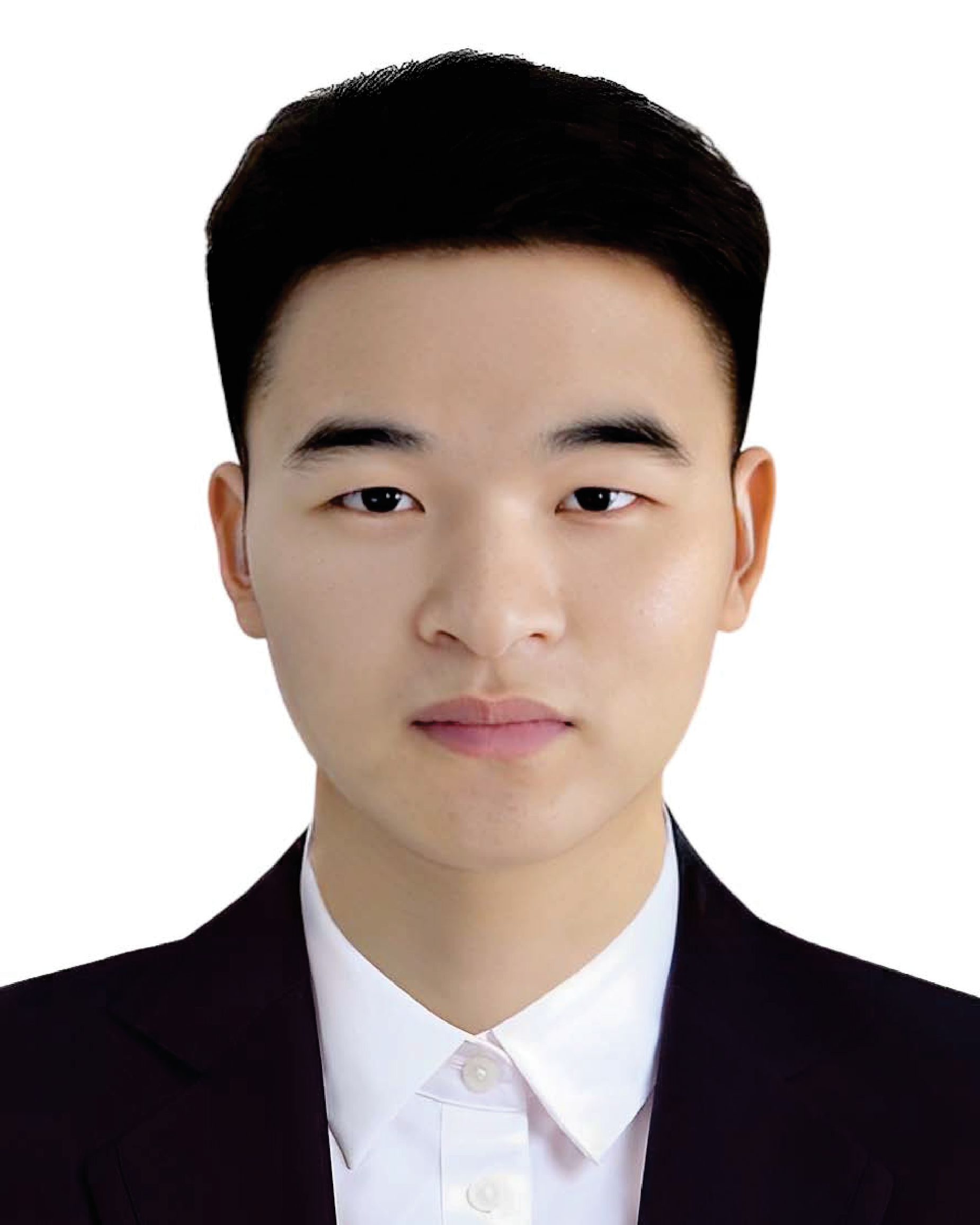}}]
	{Shuo Yuan} (Member, IEEE) received the Ph.D. degree in information and communication engineering from Beijing University of Posts and Telecommunications (BUPT), Beijing, China, in 2024.

	In 2024, he joined BUPT, where he is currently a Postdoctoral Fellow.
	His research interests include LEO satellite communication and mobile-edge computing.
	He has been a Reviewer for IEEE \textsc{Internet of Things Journal} and IEEE \textsc{Transactions on Wireless Communications}.
\end{IEEEbiography}

\vspace{-0.6in}
\begin{IEEEbiography}[{\includegraphics[width=1in,height=1.25in,clip,keepaspectratio]{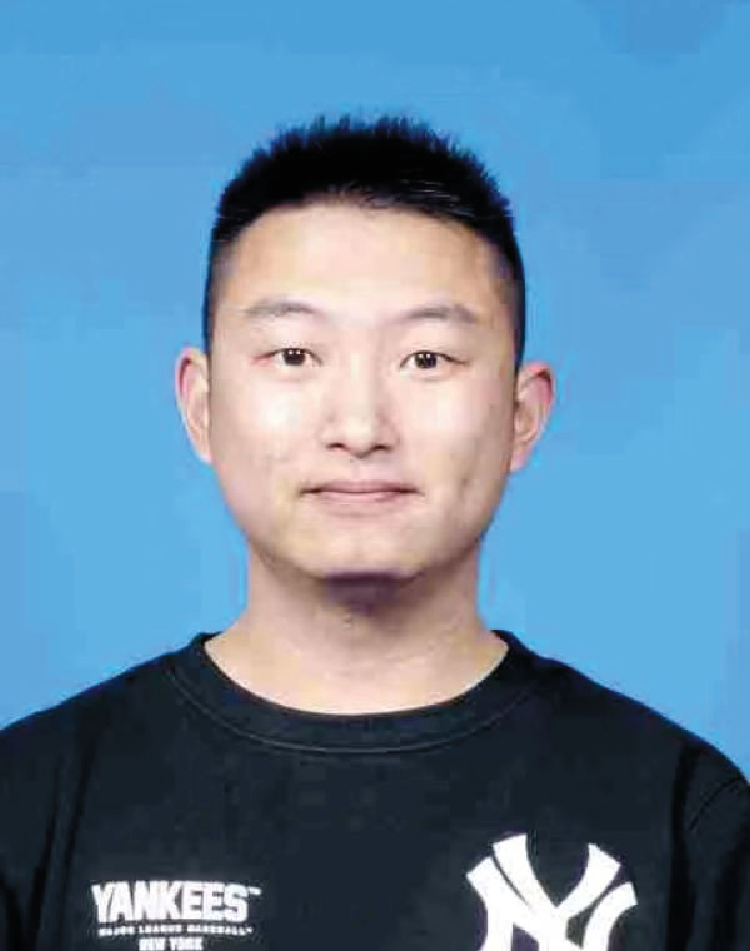}}]
	{Yaohua Sun} received the bachelor's degree (Hons.) in telecommunications engineering (with management) and the Ph.D. degree in communication engineering from Beijing University of Posts and Telecommunications (BUPT), Beijing, China, in 2014 and 2019, respectively.

	He is currently an Associate Professor with the School of Information and Communication Engineering, BUPT.
	His research interests include intelligent radio access networks and LEO satellite communication.
	He has published over 30 papers including 3 ESI highly cited papers. He has been a Reviewer for IEEE \textsc{Transactions on Communications} and IEEE \textsc{Transactions on Mobile Computing}.
\end{IEEEbiography}

\vspace{-0.6in}
\begin{IEEEbiography}[{\includegraphics[width=1in,height=1.25in,clip,keepaspectratio]{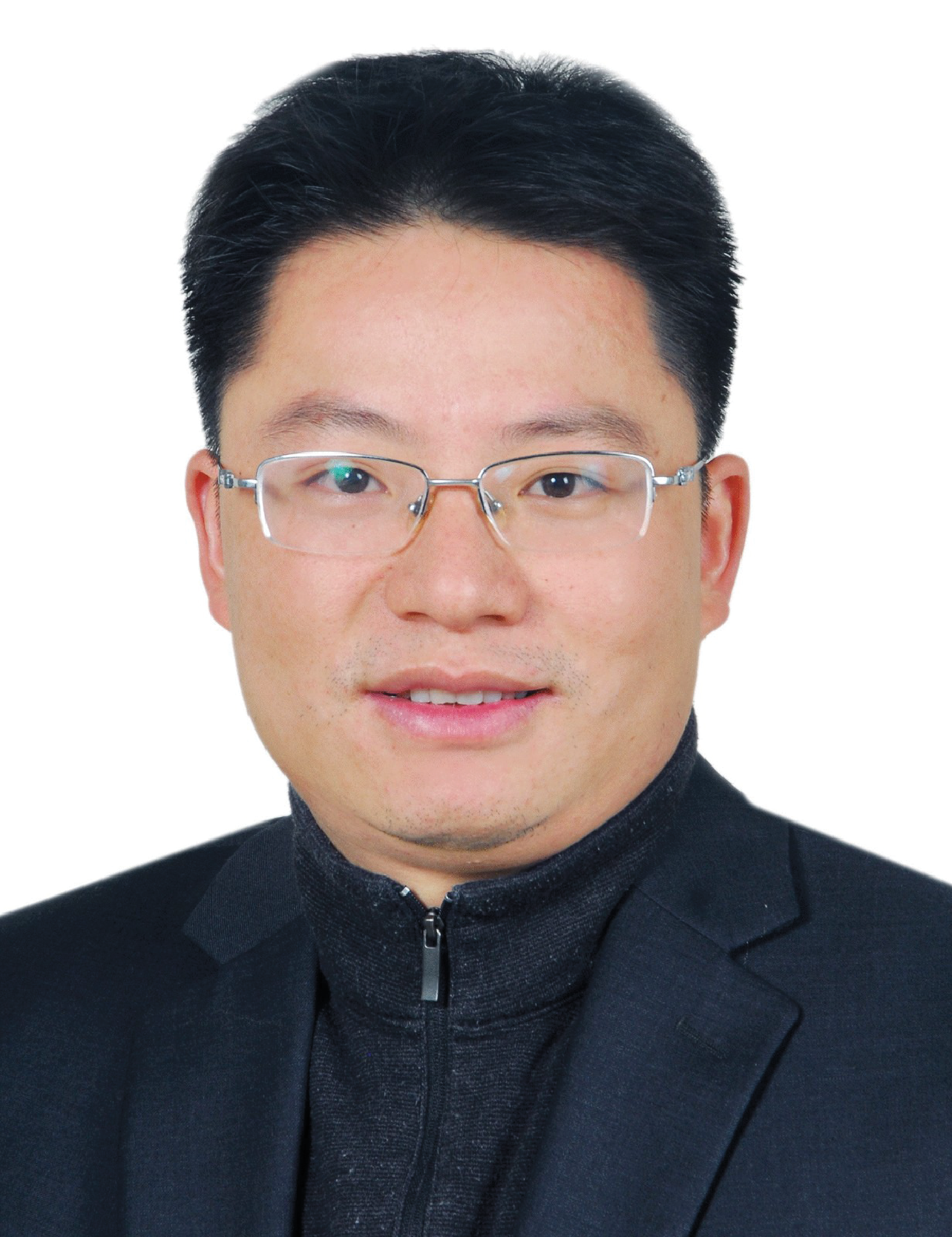}}]
	{Mugen Peng} (Fellow, IEEE) received the Ph.D. degree in communication and information systems from the Beijing University of Posts and Telecommunications, Beijing, China, in 2005. In 2014, he was an Academic Visiting Fellow at Princeton University, Princeton, NJ, USA.
	He joined BUPT, where he has been the Dean of the School of Information and Communication Engineering since June 2020, and the Deputy Director of the State Key Laboratory of Networking and Switching Technology since October 2018.
	He leads a Research Group focusing on wireless transmission and networking technologies with the State Key Laboratory of Networking and Switching Technology, BUPT.
	His main research interests include wireless communication theory, radio signal processing, cooperative communication, self-organization networking, non-terrestrial networks, and Internet of Things.
	He was a recipient of the 2018 Heinrich Hertz Prize Paper Award, the 2014 IEEE ComSoc AP Outstanding Young Researcher Award, and the Best Paper Award in IEEE/CIC ICCC 2024, IEEE ICC 2022, JCN 2016, and IEEE WCNC 2015.
	He is/was on the Editorial or Associate Editorial Board of IEEE \textsc{Communications Magazine}, IEEE \textsc{Network Magazine}, IEEE \textsc{Internet of Things Journal}, IEEE \textsc{Transactions on Vehicular Technology}, and IEEE \textsc{Transactions on Network Science and Engineering}.
\end{IEEEbiography}

\end{document}